\documentclass[11pt,onecolumn,draft]{IEEEtran}

\usepackage{epsf,psfrag,amssymb,amsfonts,cite,amsmath,eufrak,yfonts}
\newtheorem{theorem}{Theorem}
\newtheorem{example}{Example}
\newtheorem{remark}{Remark}

\newtheorem{lemma}{Lemma}
\newtheorem{definition}{Definition}

\newtheorem{proposition}{Proposition}

\def\psfancypar#1#2{\begingroup\def\par{\endgraf\endgroup\lineskiplimit=0pt}
               \setbox2=\hbox{\large\sc #2}
               \newdimen\tmpht \tmpht \ht2 \advance\tmpht by \baselineskip
               \font\hhuge=Times-Bold at \tmpht
               \setbox1=\hbox{{\hhuge #1}}
               \count7=\tmpht \count8=\ht1
               \divide\count8 by 1000 \divide\count7 by \count8
               \tmpht=.001\tmpht\multiply\tmpht by \count7
               \font\hhuge=Times-Bold at \tmpht
               \setbox1=\hbox{{\hhuge #1}}
               \noindent
                \hangindent1.05\wd1
               \hangafter=-2 {\hskip-\hangindent
               \lower1\ht1\hbox{\raise1.0\ht2\copy1}%
                \kern-0\wd1}\copy2\lineskiplimit=-1000pt}

\newenvironment{ap}{
\renewcommand{\thesection}{\mbox{Appendix }\Alph{section}}
\renewcommand{\thesubsection}{\Alph{section}.\arabic{subsection
}}
\renewcommand{\theequation}{\mbox{\Alph{section}.}\arabic{equation}}
}{
\renewcommand{\thesection}{\arabic{section}}
\renewcommand{\thesubsection}{\thesection.\arabic{subsection}}
\renewcommand{\theequation}{\arabic{equation}}
}

\newcommand{\bsq}{\begin{subequations}\begin{eqnarray}}
\newcommand{\esq}{\end{eqnarray}\end{subequations}}
\newcommand{\beq}{\begin{equation}}
\newcommand{\eeq}{\end{equation}}
\newcommand{\bqa}{\begin{eqnarray}}
\newcommand{\eqa}{\end{eqnarray}}
\newcommand{\bqn}{\begin{eqnarray*}}
\newcommand{\eqn}{\end{eqnarray*}}
\newcommand{\nn}{\nonumber}

\newcommand{\be}{\begin{enumerate}}
\newcommand{\ee}{\end{enumerate}}
\newcommand{\bi}{\begin{itemize}}
\newcommand{\ei}{\end{itemize}}
\newcommand{\bd}{\begin{description}}
\newcommand{\ed}{\end{description}}
\newcommand{\ba}{\begin{array}}
\newcommand{\ea}{\end{array}}
\newcommand{\bde}{\begin{definition}}
\newcommand{\ede}{\end{definition}}
\newcommand{\bex}{\begin{example}}
\newcommand{\eex}{\end{example}}

\newcommand{\Phibf}{\mbox{${\bf \Phi}$}}

\newcommand{\abf}{\mbox{${\bf a}$}}

\def\boxit#1{\vbox{\hrule\hbox{\vrule\kern3pt
        \vbox{\kern3pt#1\kern3pt}\kern3pt\vrule}\hrule}}

\def\reals{ { {\rm  I \kern-0.15em R }  } }
\def\complex{ {\,{{\rm C} \kern-0.50em \raise0.20ex {  |}}\, }}
\def\alphabf{\hbox{\boldmath$\alpha$\unboldmath}}

\def\gammabf{\hbox{\boldmath$\gamma$\unboldmath}}

\def\Sigmabf{\hbox{$\bf \Sigma$}}

\def\Lambdabf{\mbox{$ \bf \Lambda $}}
\def\Sigmabf{\mbox{$ \bf \Sigma $}}

\def\0bf{{\bf 0}}
\def\1bf{{\bf 1}}
\def\2bf{{\bf 2}}
\def\3bf{{\bf 3}}
\def\4bf{{\bf 4}}
\def\5bf{{\bf 5}}
\def\6bf{{\bf 6}}
\def\7bf{{\bf 7}}
\def\8bf{{\bf 8}}
\def\9bf{{\bf 9}}
\def\abf{{\bf a}}

\def\fbf{{\bf f}}

\def\hbf{{\bf h}}

\def\nbf{{\bf n}}

\def\Abf{{\bf A}}
\def\Bbf{{\bf B}}
\def\Cbf{{\bf C}}
\def\Dbf{{\bf D}}
\def\Ebf{{\bf E}}
\def\Fbf{{\bf F}}
\def\Gbf{{\bf G}}
\def\Hbf{{\bf H}}
\def\Ibf{{\bf I}}

\def\Kbf{{\bf K}}

\def\Obf{{\bf O}}

\def\Qbf{{\bf Q}}
\def\Rbf{{\bf R}}
\def\Sbf{{\bf S}}
\def\Tbf{{\bf T}}
\def\Ubf{{\bf U}}
\def\Vbf{{\bf V}}
\def\Wbf{{\bf W}}
\def\Xbf{{\bf X}}
\def\Ybf{{\bf Y}}

\def\fp{{\pmb f}}
\def\gp{{\pmb g}}
\def\hp{{\pmb h}}

\def\np{{\pmb n}}

\def\vp{{\pmb v}}

\def\xp{{\pmb x}}
\def\yp{{\pmb y}}
\def\zp{{\pmb z}}

\def\Nmat{\mathcal{N}}

\def\Smat{\mathcal{S}}

\def\Xmat{\mathcal{X}}

\def\QED{\mbox{\rule[0pt]{1.3ex}{1.3ex}}}
\def\bpf{{\em Proof: }}
\def\epf{\hspace*{\fill}~\QED\par\endtrivlist\unskip}

\def\Rxx{\Rbf_{\ssstyle X\kern-.1em X}}

\let\ssstyle=\scriptscriptstyle

\def\Cov{{\textrm{Cov}}}
\def\abs{{\textrm{abs}}}

\def\tr{{\textrm{tr}}}
\def\Vec{{\textrm{Vec}}}
\def\rank{{\textrm{rank}}}
\def\diag{{\textrm{diag}}}

\def\Kout{\setbox1=\hbox{\Huge\bf K}\hbox to
1.05\wd1{\hspace{.05\wd1}
}}
\def\Sout{\setbox1=\hbox{\Huge\bf S}\hbox to 1.05\wd1{\hspace{.05\wd1}
}}

\def\scalefig#1{\epsfxsize #1\textwidth}
 \allowdisplaybreaks[4]

\begin{document}
\title{Noisy-Interference Sum-Rate Capacity for Vector Gaussian Interference Channels}
\author{Xiaohu Shang, and H. Vincent Poor\thanks{X.
Shang is with Bell-Labs, Alcatel-Lucent, 791 Holmdel Rd., R-127, Holmdel, NJ, 07733. Email:xiaohu.shang@alcatel-lucent.com.
H. V. Poor is with Princeton University, Department of Electrical Engineering, Princeton, NJ, 08544. Email: poor@princeton.edu.
H. V. Poor was supported in part by the National
Science Foundation under Grant CNS-09-05398.}} \maketitle

\begin{abstract}
New sufficient conditions for a vector Gaussian interference channel to achieve the sum-rate capacity by treating interference as noise are derived, which generalize the existing results. More concise conditions for multiple-input-single-output, and single-input-multiple-output scenarios are obtained.
\end{abstract}


\section{Introduction}
The interference channel (IC) was first introduce by Shannon \cite{Shannon:61Berkeley}, and was later studied by Ahlswede \cite{Ahlswede:74AP} who gave a limiting expression for the capacity region. Determination of the single-letter expression of the capacity region of an IC has been a long standing open problem ever since.

The first capacity region of the IC was obtained by Carleial in \cite{Carleial:75IT} for the very strong interference case, in which the capacity is achieved by decoding and subtracting the interference before decoding the useful signals. The Gaussian IC model with power constraint was also introduced in \cite{Carleial:75IT}. The result of \cite{Carleial:75IT} was later extended to discrete memoryless ICs in \cite{Sato&Tanabe:78IECE}. In \cite{Carleial:78IT}, Carleial showed that any Gaussian IC can be written in the standard form, i.e., both direct links have unit channel gain and the Gaussian noise has unit variance. An inner bound on the capacity region was obtained in \cite{Carleial:78IT} using superposition coding and sequential decoding. The best inner bound was obtained in \cite{Han&Kobayashi:81IT} using superposition coding and joint decoding. This inner bound was later simplified in \cite{Chong-etal:08IT} and \cite{Kramer:06Zurich}. Early outer bounds on the capacity region of the IC can be found in \cite{Sato:77IT}, \cite{Sato:78IT} and \cite{Carleial:83IT}. The capacity region of Gaussian IC with strong interference was obtained in \cite{Han&Kobayashi:81IT} and \cite{Sato:81IT}, in which jointly decoding both the interference and the useful signal achieves the capacity. This result was extended to discrete memoryless ICs in \cite{Costa&ElGamal:87IT}. The degraded memoryless IC was studied in \cite{Benzel:79IT} and later in \cite{Liu&Ulukus:08IT}. The degraded Gaussian IC was studied in \cite{Sato:81IT} and the sum-rate capacity was obtained. It was shown in \cite{Costa:85IT} that the capacity region of a Gaussian Z interference channel (ZIC) is equivalent to that of a degraded Gaussian IC. Therefore, the sum-rate capacity of a Gaussian ZIC is automatically obtained. The corner points of the capacity region of a Gaussian IC were also studied in \cite{Costa:85IT} and this still remains an open problem \cite{Sason:04IT}. In \cite{Cheng&Verdu:93IT}, it has been shown that Gaussian inputs do not achieve the capacity region of the Gaussian IC in the limiting expression of \cite{Ahlswede:74AP}.

In \cite{Kramer:04IT}, two outer bounds on the capacity region were derived. The first bound is based on a genie-aided approach in which additional information is provided to the receivers. The second bound of \cite{Kramer:04IT} is obtained by allowing cooperation between transmitters. It was speculated in \cite{Kramer:04IT} that there might be other genies which give tighter outer bound than \cite[Theorem 1]{Kramer:04IT}. In \cite{Etkin-etal:08IT} another outer bound was derived using different genies. Using this bound, the Han and Kobayashi inner bound \cite{Han&Kobayashi:81IT} is shown to be within $1$ bit of the capacity region. Motivated  by \cite{Etkin-etal:08IT}, new outer bounds were derived in \cite{Shang-etal:09IT,Motahari&Khandani:09IT,Annapureddy&Veeravalli:09IT} and it was shown that the sum-rate capacity is achieved by treating interference as noise if the IC satisfies a simple condition. This kind of Gaussian IC is said to have noisy interference. This noisy-interference sum-rate capacity is extended to multi-user Gaussian ICs in \cite{Shang-etal:08ISIT,Shang-etal:08Milcom,Annapureddy&Veeravalli:09IT}. Meanwhile, the sum-rate capacity for Gaussian ICs with mix-interference was determined in \cite{Motahari&Khandani:09IT} and \cite{Weng&Tuninetti:08ITA} using \cite[Theorem 1]{Kramer:04IT}.

\begin{figure}[h] \centerline{
\begin{psfrags}
\psfrag{x1}[c]{$\xp_1$} \psfrag{x2}[c]{$\xp_2$}
\psfrag{y1}[c]{$\yp_1$} \psfrag{y2}[c]{$\yp_2$}
\psfrag{n1}[c]{$\zp_1$} \psfrag{n2}[c]{$\zp_2$} \psfrag{+}[c]{$+$}
\psfrag{g11}[c]{$\Hbf_1$} \psfrag{g12}[c]{$\Fbf_1$}
\psfrag{g21}[c]{$\Fbf_2$} \psfrag{g22}[c]{$\Hbf_2$}
\scalefig{.35}\epsfbox{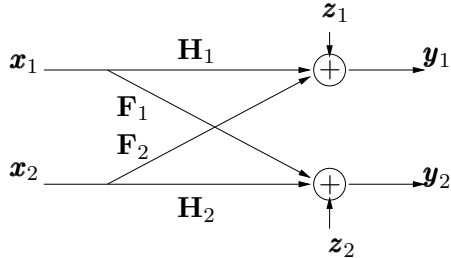}
\end{psfrags}}
\caption{\label{fig:model} The two-user MIMO IC.}
\end{figure}

In this paper, we study the capacity of the two-user
multiple-input multiple-output (MIMO) IC. As shown in Fig.
\ref{fig:model}, the received signals are defined as
 \bqa
&&\hspace{-.2in}\yp_1=\Hbf_1\xp_1+\Fbf_2\xp_2+\zp_1\nn\\
&&\hspace{-.2in}\yp_2=\Hbf_2\xp_2+\Fbf_1\xp_1+\zp_2%
\label{eq:model}%
\eqa
where $\xp_i,i=1,2,$ is the transmitted (column) vector signal of user $i$ which
is subject to the average power constraint
\bqa%
\sum_{j=1}^n\tr\left(E\left[\xp_{ij}\xp_{ij}^T\right]\right)\leq nP_i
\label{eq:CovConstraint}
\eqa%
where $\xp_{i1},\xp_{i2},\ldots,\xp_{in}$, is the transmitted vector
sequence of user $i$, and $P_i$ is the power constraint.
The noise $\zp_i$ is a
Gaussian random vector with zero mean and identity covariance
matrix; and $\Hbf_{i}$ and $\Fbf_i$, $i=1,2$, are the channel
matrices known at both the transmitters and receivers. Transmitter $i$ has $t_i$ antennas and receiver $i$ has $r_i$ antennas. Without loss of generality, we assume $\Hbf_i\neq\0bf$ and $P_i>0$.

The capacity of a MIMO IC was first studied in \cite{Vishwanath&Jafar:04ITW} which derived an outer bound on the capacity region and determined the capacity region for the single-input-multiple-output (SIMO) IC with strong interference. A lower bound for the sum-rate capacity based on Han and
Kobayashi's region were discussed in \cite{Shang-etal:06IT}. Telatar and Tse \cite{Telatar&Tse:07ISIT} showed
that Han and Kobayashi's region is within one bit per receive
antenna of the capacity region.  Recent work in \cite{Shang-etal:08Allerton} and \cite{Shang-etal:10IT_mimo} extended
the existing capacity results from scalar ICs to MIMO ICs under
average power constraints. Specifically, \cite{Shang-etal:08Allerton} and \cite{Shang-etal:10IT_mimo} derived the capacity
region for aligned-strong interference, and the sum-rate capacity for aligned-strong Z
interference, aligned-weak Z interference, noisy interference and mixed interference under
average power constraints. In \cite{Shang-etal:10IT_mimo}, we say that a MIMO IC has
\bi
\item aligned-strong interference if $\Hbf_i=\Fbf_i\Abf_i$, $i=1,2$; or aligned strong Z interference: if $\Fbf_1=\0bf$ and $\Hbf_2=\Fbf_2\Abf_2$;
\item aligned-weak Z interference: if $\Fbf_1=\0bf$ and $\Fbf_2=\Hbf_2\Abf_2$;
\item noisy interference if \cite[(36)-(39)]{Shang-etal:10IT_mimo} are satisfied for all $\Sbf_i\succeq\0bf$ with $\tr(\Sbf_i)\leq P_i$; and
\item mixed interference if $\Hbf_1=\Fbf_1\Abf_1$ and $\Fbf_2=\Hbf_2\Abf_2$;
\ei
where $\Abf_i$ is a matrix satisfying $\Abf_i\Abf_i^T\preceq\Ibf$, and $\Ibf$ is an identity matrix. It can be shown that the capacity region of the SIMO IC with strong interference \cite{Vishwanath&Jafar:04ITW} is a special case of that of the aligned-strong interference. Moreover, the capacity results for aligned-strong interference, aligned-strong or aligned-weak Z interference and mixed-interference apply to other power constraints, e.g., a covariance matrix constraint, a peak power constraint and a per-antenna power constraint.

The noisy-interference condition for MIMO ICs was later studied in \cite{Annapureddy&Veeravalli:09IT_submission} which requires only the optimal covariance matrices of $\xp_1$ and $\xp_2$ to satisfy the conditions \cite[(36)-(39)]{Shang-etal:10IT_mimo}, as long as these
optimal covariance matrices are of full rank. An application of this result is the noisy-interference sum-rate capacity for symmetric SIMO ICs, i.e., $\Hbf_i$ and $\Fbf_i$ are column vectors with $\Hbf_1=\Hbf_2$ and $\Fbf_1=\Fbf_2$ and the power constraints are identical $P_1=P_2$.

The results of \cite{Shang-etal:10IT_mimo} and \cite{Annapureddy&Veeravalli:09IT_submission} on the MIMO IC with noisy interference obtain different power regions. Intuitively, \cite{Shang-etal:10IT_mimo} obtains the low power region of the noisy interference and \cite{Annapureddy&Veeravalli:09IT_submission} obtains the comparatively high power region of the noisy interference. The reason is that, \cite{Shang-etal:10IT_mimo} requires the power to be low enough such that any power allocation satisfies conditions \cite[(36)-(39)]{Shang-etal:10IT_mimo}; while \cite{Annapureddy&Veeravalli:09IT_submission} requires the power to be high enough such that each eigen-mode is allocated non-zero power, and \cite[(36)-(39)]{Shang-etal:10IT_mimo} are satisfied.

There exist MIMO ICs with noisy interference but which are not in the categories of \cite{Shang-etal:10IT_mimo} or \cite{Annapureddy&Veeravalli:09IT_submission}. These MIMO ICs include the parallel Gaussian IC \cite{Shang-etal:11IT_pgic} in which $\Hbf_i$ and $\Fbf_i$ are diagonal matrices, and the symmetric multiple-input-single-output (MISO) IC \cite{Annapureddy&Veeravalli:09IT_submission} in which $\Hbf_i$ and $\Fbf_i$ are row vectors with $\Hbf_1=\Hbf_2$ and $\Fbf_1=\Fbf_2$ and the power constraints are identical $P_1=P_2$. For the noisy-interference conditions of both the parallel Gaussian IC and the symmetric MISO IC, there may exist some power allocations that violate \cite[(36)-(39)]{Shang-etal:10IT_mimo}. Furthermore, the optimal input covariance matrices for the parallel Gaussian IC can be singular, and the optimal input covariance matrices for the symmetric MISO IC is always rank-$1$. Therefore, neither \cite{Shang-etal:10IT_mimo} nor \cite{Annapureddy&Veeravalli:09IT_submission} applies to these two special cases.

The major difficulty in the determination of the noisy-interference sum-rate capacity of a MIMO IC is that the characterization of the optimal input covariance matrices by treating interference as noise is needed in the derivation. However, these optimal input covariance matrices are unknown due to the non-convex nature of the optimization problem for maximizing the sum rate of single-user detection. In \cite{Shang-etal:10IT_mimo} all the possible input covariance matrices are required to satisfy some conditions. The results in \cite{Annapureddy&Veeravalli:09IT_submission} and \cite{Shang-etal:11IT_pgic}, although not requiring all the input covariance matrices to satisfy the conditions, they do have some assumptions, or have some knowledge on the optimal input covariance matrices:
\bi
\item Special MIMO ICs in \cite{Annapureddy&Veeravalli:09IT_submission}: the optimal input covariance matrices are assumed to be of full rank.
\item Parallel Gaussian IC in \cite{Shang-etal:11IT_pgic}: the optimal input covariance matrices are diagonal. More importantly, the optimal power allocated at each antenna satisfies the parallel supporting hyperplane condition, or in another words, the sum-rate function for each sub-channel has the same subgradient at the optimal power allocation.
\item Symmetric MISO IC in \cite{Annapureddy&Veeravalli:09IT_submission}: beamforming achieves the largest sum-rate for treating interference as noise. Thus the optimal input covariance matrices are both rank-$1$. The optimality of beamforming was proved in \cite{Shang&Chen:07Asi} and \cite{Shang-etal:09IT_submission2}. The same result was reproduced using different methods in \cite{Zhang&Cui:10SP} and \cite{Mochaourab&Jorswieck:10SP_submission}. By restricting to rank-$1$ matrices and using the assumption that the MISO IC is symmetric, the closed-form optimal input covariance matrices are obtained, which is crucial in deriving the noisy interference condition.
\ei

In this paper, we revisit the sum-rate capacity of the MIMO IC and derive a new noisy-interference condition, i.e., treating interference as noise achieves the sum-rate capacity. This new condition requires only the optimal input covariance matrices to satisfy \cite[(36)-(39)]{Shang-etal:10IT_mimo} and an additional condition, but does not require the optimal input covariance matrices to be of full rank (when they are of full rank, this additional condition is automatically satisfied). Thus, this new noisy-interference condition includes those in \cite{Shang-etal:10IT_mimo} and \cite{Annapureddy&Veeravalli:09IT_submission} as special cases. In addition, this noisy-interference condition includes those of the parallel Gaussian IC \cite{Shang-etal:11IT_pgic} and the symmetric MISO IC \cite{Annapureddy&Veeravalli:09IT_submission} as special cases. More concise condition for the general asymmetric MISO or SIMO ICs are also obtained.

The rest of the paper is organized as follows: the noisy-interference sum-rate capacity for the MIMO IC is obtained in Section \ref{section:MIMO}; the MISO and SIMO ICs are discussed in Sections \ref{section:MISO} and \ref{section:SIMO}, respectively; numerical examples are given in Section \ref{section:example}; and we conclude in Section \ref{section:conclusion}.

Before proceeding, we introduce some notation that will be used in
the paper.
\bi %
\item Italic letters (e.g. $X$) denote scalars; and
bold letters $\xp$ and $\Xbf$ denote column vectors and matrices,
respectively.
\item $\Ibf$ denotes the identity matrix
and $\0bf$ denotes the all-zero vector or matrix. The dimensions of $\Ibf$ and $\0bf$ are determined by the context.
\item $|\Xbf|$, $\Xbf^T$, $\Xbf^{-1}$ and $\rank(\Xbf)$ denote
respectively the determinant, transpose,
inverse, and rank of the matrix $\Xbf$, and $\|\xp\|$ denotes the Euclidean vector norm of $\xp$, i.e., $\|\xp\|^2=\xp^T\xp$.
\item $\textrm{radius}(\Xbf)$ is the numerical radius \cite[p. 321]{Horn&Johnson:book} of the square real matrix $\Xbf$, and is defined as
\bqn
\textrm{radius}(\Xbf)=\max_{\alphabf^T\alphabf\leq 1}\abs\left(\alphabf^T\Xbf\alphabf\right),
\eqn
where $\alphabf$ is a vector, and $\abs(\cdot)$ denotes the absolute value.
\item
$\xp^n=\left[\xp_1^T,\xp_2^T,\dots,\xp_n^T\right]^T$ is a long
vector that consists of a sequence of vectors $\xp_i, i=1,\dots,
n$. $\diag[X_1,\cdots,X_n]$ is a diagonal matrix with diagonal entries $X_i$.
\item $\Vec\left(\Abf\right)$ denote the vectorization operator, i.e., let $\Abf=[\abf_1,\abf_2,\cdots,\abf_n]$, and $\abf_i,i=1,\cdots,n$ be the column vectors, then $\Vec\left(\Abf\right)=[\abf_1^T,\abf_2^T,\cdots,\abf_n^T]^T$.
\item
$\xp\sim\Nmat\left(\0bf,\Sigmabf\right)$ means that the random
vector $\xp$ has Gaussian distribution with zero mean and covariance
matrix $\Sigmabf$.
\item $E[\cdot]$ denotes expectation;
$\textrm{Cov}(\cdot)$ denotes covariance matrix; $I(\cdot;\cdot)$
denotes mutual information; $h(\cdot)$ denotes differential
entropy with the logarithm base $e$, and
$\log(\cdot)=\log_e(\cdot)$.

\ei

\section{MIMO ICs}
\label{section:MIMO}

We first derive a lower bound and an upper bound on the sum-rate capacity. The lower bound is simply the single-user detection sum rate. The upper bound is obtained by providing the receivers with appropriate side information. Both the lower and upper bounds are formulated as optimization problems in which the lower bound is a non-convex problem and the upper bound is a convex problem. The sum-rate capacity is obtained by determining conditions under which these two optimization problems have the same solution.

\subsection{Lower bound on the sum-rate capacity}

By treating interference as noise, the maximum of the following optimization problem is a lower bound on the sum-rate capacity:
\bqa
\max&&\quad \frac{1}{2}\log\left|\Ibf+\Hbf_1\Sbf_1\Hbf_1^T\left(\Ibf+\Fbf_2\Sbf_2\Fbf_2^T\right)^{-1}\right|+\frac{1}{2}\log\left|\Ibf+\Hbf_2\Sbf_2\Hbf_2^T\left(\Ibf+\Fbf_1\Sbf_1\Fbf_1^T\right)^{-1}\right|\nn\\
\textrm{subject to}&&\quad \tr(\Sbf_1)\leq P_1,\quad \tr(\Sbf_2)\leq P_2\nn\\
&&\quad \Sbf_1\succeq\0bf,\quad \Sbf_2\succeq\0bf.
\label{eq:lowerMIMO}
\eqa

The following lemma gives the necessary Karush-Kuhn-Tucker (KKT) conditions for the optimal input covariance matrices $\Sbf_i^*$, $i=1,2$.
\begin{lemma}
Let $\Sbf_1^*$ and $\Sbf_2^*$ be optimal for problem (\ref{eq:lowerMIMO}), if $P_1,P_2>0$, then there exist scalars $\lambda_i$ and matrices $\Wbf_i$, $i=1,2$, such that
\bqa
&&\Gbf_1+\lambda_1\Ibf-\Wbf_1=\0bf
\label{eq:lowerKKTS1}\\
&&\Gbf_2+\lambda_2\Ibf-\Wbf_2=\0bf
\label{eq:lowerKKTS2}\\
&&\lambda_i\left\{\begin{array}{ll}
                     >0 &\quad \textrm{if } \tr(\Sbf_i^*)=P_i \\
                     =0 &\quad \textrm{if } \tr(\Sbf_i^*)<P_i
                   \end{array}
 \right.\quad i=1,2
\label{eq:lowerEq}\\
&&\tr(\Sbf_i^*\Wbf_i)=0,\quad \Wbf_i\succeq\0bf \quad i=1,2
\label{eq:lowerIneq}
\eqa
where
\bqa
\Gbf_1&{}={}&-\left.\frac{\partial R_{1l}}{\partial\Sbf_1}\right|_{\Sbf_i=\Sbf_i^*}-\left.\frac{\partial R_{2l}}{\partial\Sbf_1}\right|_{\Sbf_i=\Sbf_i^*}
\label{eq:G1}\\
\Gbf_2&{}={}&-\left.\frac{\partial R_{1l}}{\partial\Sbf_2}\right|_{\Sbf_i=\Sbf_i^*}-\left.\frac{\partial R_{2l}}{\partial\Sbf_2}\right|_{\Sbf_i=\Sbf_i^*}
\label{eq:G2}\\
\left.\frac{\partial R_{1l}}{\partial\Sbf_1}\right|_{\Sbf_i=\Sbf_i^*}&{}={}&\frac{1}{2}\Hbf_1^T\left(\Ibf+\Hbf_1\Sbf_1^*\Hbf_1^T+\Fbf_2\Sbf_2^*\Fbf_2^T\right)^{-1}\Hbf_1
\\
\left.\frac{\partial R_{1l}}{\partial\Sbf_2}\right|_{\Sbf_i=\Sbf_i^*}&{}={}&-\frac{1}{2}\Fbf_2^T\left[\left(\Ibf+\Fbf_2\Sbf_2^*\Fbf_2^T\right)^{-1}-\left(\Ibf+\Hbf_1\Sbf_1^*\Hbf_1^T+\Fbf_2\Sbf_2^*\Fbf_2^T\right)^{-1}\right]\Fbf_2\\
\left.\frac{\partial R_{2l}}{\partial\Sbf}\right|_{\Sbf_1=\Sbf_i^*}&{}={}&-\frac{1}{2}\Fbf_1^T\left[\left(\Ibf+\Fbf_1\Sbf_1^*\Fbf_1^T\right)^{-1}-\left(\Ibf+\Hbf_2\Sbf_2^*\Hbf_2^T+\Fbf_1\Sbf_1^*\Fbf_1^T\right)^{-1}\right]\Fbf_1\\
\left.\frac{\partial R_{2l}}{\partial\Sbf_2}\right|_{\Sbf_i=\Sbf_i^*}&{}={}&\frac{1}{2}\Hbf_2^T\left(\Ibf+\Hbf_2\Sbf_2^*\Hbf_2^T+\Fbf_1\Sbf_1^*\Fbf_1^T\right)^{-1}\Hbf_2
\eqa
and
\bqa
R_{1l}\left(\Sbf_1,\Sbf_2\right)&{}={}&\frac{1}{2}\log\left|\Ibf+\Hbf_1\Sbf_1\Hbf_1^T\left(\Ibf+\Fbf_2\Sbf_2\Fbf_2^T\right)^{-1}\right|
\label{eq:R1l}\\
R_{2l}\left(\Sbf_1,\Sbf_2\right)&{}={}&\frac{1}{2}\log\left|\Ibf+\Hbf_2\Sbf_2\Hbf_2^T\left(\Ibf+\Fbf_1\Sbf_1\Fbf_1^T\right)^{-1}\right|.
\label{eq:R2l}
\eqa
\label{lemma:KKT}
\end{lemma}
\bpf
Conditions (\ref{eq:lowerKKTS1})-(\ref{eq:lowerIneq}) are the KKT conditions for problem (\ref{eq:lowerMIMO}).
Here, we only need to prove that problem (\ref{eq:lowerMIMO}) satisfies some constraint qualifications denoted by CQ5 in \cite[p. 306]{Bertsekas-etal:book} such that $\lambda_i$ and $\Wbf_i$ do exist. The rest of the proof is included in Appendix \ref{appendix:kkt}.
\epf

\subsection{Upper bound on the sum-rate capacity}

The following is an upper bound on the sum-rate capacity of a MIMO IC.

\begin{theorem}
The sum-rate capacity of the MIMO IC is upper bounded by the maximum achieved in the following optimization problem:
\bqa
\max&&\quad \frac{1}{2}\log\left|\Ibf+\left[\begin{array}{c}
                                                    \Hbf_1 \\
                                                    \Fbf_1
                                                  \end{array}
\right]\Sbf_1\left[\begin{array}{c}
                                                    \Hbf_1 \\
                                                    \Fbf_1
                                                  \end{array}
\right]^T\left(\Ebf_1+\left[\begin{array}{c}
                            \Fbf_2 \\
                            \0bf
                          \end{array}
\right]\Sbf_2\left[\begin{array}{c}
                            \Fbf_2 \\
                            \0bf
                          \end{array}
\right]^T\right)^{-1}\right|\nn\\
&&\qquad+\frac{1}{2}\log\left|\Ibf+\left[\begin{array}{c}
                                                    \Hbf_2 \\
                                                    \Fbf_2
                                                  \end{array}
\right]\Sbf_2\left[\begin{array}{c}
                                                    \Hbf_2 \\
                                                    \Fbf_2
                                                  \end{array}
\right]^T\left(\Ebf_2+\left[\begin{array}{c}
                            \Fbf_1 \\
                            \0bf
                          \end{array}
\right]\Sbf_1\left[\begin{array}{c}
                            \Fbf_1 \\
                            \0bf
                          \end{array}
\right]^T\right)^{-1}\right|\nn\\
\textrm{subject to}&&\quad \tr(\Sbf_1)\leq P_1,\quad \tr(\Sbf_2)\leq P_2\nn\\
&&\quad \Sbf_1\succeq\0bf,\quad \Sbf_2\succeq\0bf
\label{eq:upper}
\eqa
where $\Ebf_i, i=1,2$, can be any symmetric positive definite matrix satisfying
\bqa
\Ebf_i&{}={}&\left[\begin{array}{cc}
               \Ibf &\quad \Abf_i \\
               \Abf_i^T &\quad \Sigmabf_i
             \end{array}
\right]\succ\0bf
\label{eq:Ei}\\
\Sigmabf_1&{}\preceq{}&\Ibf-\Abf_2\Sigmabf_2^{-1}\Abf_2^T
\label{eq:sigma1}\\
\Sigmabf_2&{}\preceq{}&\Ibf-\Abf_1\Sigmabf_1^{-1}\Abf_1^T.
\label{eq:sigma2}
\eqa
\label{theorem:upper}
\end{theorem}
\bpf
Let $\nbf_i^n, i=1,2,$ be a length-$n$ sequence of independent and identically distributed (i.i.d.) Gaussian vectors, each having joint distribution with $\zp_i$ given by
\bqa
\left[\begin{array}{c}
        \zp_i \\
        \np_i
      \end{array}
\right]\sim\Nmat\left(\0bf,\Ebf_i\right)=\Nmat\left(\0bf,\left[\begin{array}{cc}
               \Ibf &\quad \Abf_i \\
               \Abf_i^T &\quad \Sigmabf_i
             \end{array}
\right]\right).
\label{eq:ZN}
\eqa
Let $\xp_i^n$ be the input sequence of user $i$, and
\bqa
&&\sum_{j=1}^n\Cov\left(\xp_{ij}\right)=n\Sbf_i\\
&&\tr\left(\Sbf_i\right)\leq P_i
\eqa
Let $\epsilon>0$ and $\epsilon\rightarrow 0$ when $n\rightarrow\infty$. Then for any achievable rate $R_1$ and $R_2$, we have
\bqa
&&\hspace{-.1in}n(R_1+R_2)-n\epsilon\nn\\
&&\hspace{-.1in}\leq I\left(\xp_1^n;\Hbf_1\xp_1^n+\Fbf_2\xp_2^n+\zp_1^n\right)+I\left(\xp_2^n;\Hbf_2\xp_2^n+\Fbf_1\xp_1^n+\zp_2^n\right)\nn\\
&&\hspace{-.1in}\leq I\left(\xp_1^n;\Hbf_1\xp_1^n+\Fbf_2\xp_2^n+\zp_1^n,\Fbf_1\xp_1^n+\np_1^n\right)+I\left(\xp_2^n;\Hbf_2\xp_2^n+\Fbf_1\xp_1^n+\zp_2^n,\Fbf_2\xp_2^n+\np_2^n\right)\nn\\
&&\hspace{-.1in}= h\left(\Fbf_1\xp_1^n+\np_1^n\right)-h\left(\np_1^n\right)+h\left(\Hbf_1\xp_1^n+\Fbf_2\xp_2^n+\zp_1^n|\Fbf_1\xp_1^n+\np_1^n\right)-h\left(\Fbf_2\xp_2^n+\zp_1^n|\np_1^n\right)\nn\\
  &&\hspace{-.1in}\hspace{.2in}+h\left(\Fbf_2\xp_2^n+\np_2^n\right)-h\left(\np_2^n\right)+h\left(\Hbf_2\xp_2^n+\Fbf_1\xp_1^n+\zp_2^n|\Fbf_2\xp_2^n+\np_2^n\right)-h\left(\Fbf_1\xp_1^n+\zp_2^n|\np_2^n\right)\nn\\
&&\hspace{-.1in}\stackrel{(a)}\leq h\left(\Fbf_1\xp_1^n+\np_1^n\right)-nh\left(\np_1\right)+nh\left(\Hbf_1\xp_{1G}+\Fbf_2\xp_{2G}+\zp_1|\Fbf_1\xp_{1G}+\np_1\right)-h\left(\Fbf_2\xp_2^n+\zp_1^n|\np_1^n\right)\nn\\
  &&\hspace{-.1in}\hspace{.2in}+h\left(\Fbf_2\xp_2^n+\np_2^n\right)-nh\left(\np_2\right)+nh\left(\Hbf_2\xp_{2G}+\Fbf_1\xp_{1G}+\zp_2|\Fbf_2\xp_{2G}+\np_2\right)-h\left(\Fbf_1\xp_1^n+\zp_2^n|\np_2^n\right)\nn\\
&&\hspace{-.1in}\stackrel{(b)}\leq nh\left(\Fbf_1\xp_{1G}+\np_1\right)-nh\left(\np_1\right)+nh\left(\Hbf_1\xp_{1G}+\Fbf_2\xp_{2G}+\zp_1|\Fbf_1\xp_{1G}+\np_1\right)-nh\left(\Fbf_2\xp_{2G}+\zp_1|\np_1\right)\nn\\
  &&\hspace{-.1in}\hspace{.2in}+nh\left(\Fbf_2\xp_{2G}+\np_2\right)-nh\left(\np_2\right)+nh\left(\Hbf_2\xp_{2G}+\Fbf_1\xp_{1G}+\zp_2|\Fbf_2\xp_{2G}+\np_2\right)-nh\left(\Fbf_1\xp_{1G}+\zp_2|\np_2\right)
\label{eq:concaveExp}\\
&&\hspace{-.1in}=nI\left(\xp_{1G};\left[\begin{array}{c}
                                                    \Hbf_1 \\
                                                    \Fbf_1
                                                  \end{array}
\right]\xp_{1G}+\left[\begin{array}{c}
                                                    \Fbf_2 \\
                                                    \0bf
                                                  \end{array}
\right]\xp_{2G}+\left[\begin{array}{c}
                                                    \zp_1 \\
                                                    \np_1
                                                  \end{array}
\right]\right)+nI\left(\xp_{1G};\left[\begin{array}{c}
                                                    \Hbf_2 \\
                                                    \Fbf_2
                                                  \end{array}
\right]\xp_{2G}+\left[\begin{array}{c}
                                                    \Fbf_1 \\
                                                    \0bf
                                                  \end{array}
\right]\xp_{1G}+\left[\begin{array}{c}
                                                    \zp_2 \\
                                                    \np_2
                                                  \end{array}
\right]\right)
\label{eq:upperMutualI}
\eqa
where in (a) we define $\xp_{iG}\sim\Nmat\left(\0bf,\Sbf_i\right)$ and the inequality is by \cite[Lemma 2]{Shang-etal:10IT_mimo}, and (b) is by (\ref{eq:sigma1}), (\ref{eq:sigma2}) and \cite[Lemma 3]{Shang-etal:10IT_mimo}.
\epf

The following lemma establishes the convexity of the optimization problem (\ref{eq:upper}) and the proof is included in Appendix \ref{appendix:cvx}.
\begin{lemma}
The optimization problem (\ref{eq:upper}) is a convex optimization problem.
\label{lemma:cvxOpt}
\end{lemma}

Theorem \ref{theorem:upper} is derived using the same method that has been used in \cite{Shang-etal:10IT_mimo}. The maximum achieved in problem (\ref{eq:upper}) for any choice of $\Abf_i$ and $\Sigmabf_i$ that satisfy (\ref{eq:Ei})-(\ref{eq:sigma2}) is an upper bound on the sum-rate capacity of this MIMO IC regardless of whether it has noisy interference or not.

\subsection{Sum-rate capacity}

When the MIMO IC has noisy interference, we can choose appropriate $\Abf_i$ and $\Sigmabf_i$ such that the lower and upper bounds converge. Before proceeding, we first introduce the following matrix identity which will be used repeatedly in the proof of our main result.
\begin{lemma} Assuming all the matrices have feasible dimension and the relevant matrices are invertible, we have
\bqa
\left[\begin{array}{cc}
        \Abf_{11} &\quad \Abf_{12} \\
        \Abf_{21} &\quad \Abf_{22}
      \end{array}
\right]^{-1}=\left[\begin{array}{cc}
        \Abf_{11}^{-1} &\quad \0bf \\
        \0bf &\quad \0bf
      \end{array}
\right]+\left[\begin{array}{c}
                \Abf_{11}^{-1}\Abf_{12} \\
                -\Ibf
              \end{array}
\right]\left(\Abf_{22}-\Abf_{21}\Abf_{11}^{-1}\Abf_{12}\right)^{-1}\left[\begin{array}{cc}
                \Abf_{21}\Abf_{11}^{-1} &\quad -\Ibf
              \end{array}
\right].
\eqa
\label{lemma:myMxInv}
\end{lemma}
\bpf
\bqa
&&\left[\begin{array}{cc}
        \Abf_{11} &\quad \Abf_{12} \\
        \Abf_{21} &\quad \Abf_{22}
      \end{array}
\right]^{-1}\nn\\
&&\stackrel{(a)}=\left[\begin{array}{cc}
           \left(\Abf_{11}-\Abf_{12}\Abf_{22}^{-1}\Abf_{21}\right)^{-1} &\quad -\Abf_{11}^{-1}\Abf_{12}\left(\Abf_{22}-\Abf_{21}\Abf_{11}^{-1}\Abf_{12}\right)^{-1} \\
           -\left(\Abf_{22}-\Abf_{21}\Abf_{11}^{-1}\Abf_{12}\right)^{-1}\Abf_{21}\Abf_{11}^{-1} &\quad \left(\Abf_{22}-\Abf_{21}\Abf_{11}^{-1}\Abf_{12}\right)^{-1}
         \end{array}
\right]\nn\\
&&\stackrel{(b)}=\left[\begin{array}{cc}
           \Abf_{11}^{-1}+\Abf_{11}^{-1}\Abf_{12}\left(\Abf_{22}-\Abf_{21}\Abf_{11}^{-1}\Abf_{12}\right)^{-1}\Abf_{21}\Abf_{11}^{-1} &\quad -\Abf_{11}^{-1}\Abf_{12}\left(\Abf_{22}-\Abf_{21}\Abf_{11}^{-1}\Abf_{12}\right)^{-1} \\
           -\left(\Abf_{22}-\Abf_{21}\Abf_{11}^{-1}\Abf_{12}\right)^{-1}\Abf_{21}\Abf_{11}^{-1} &\quad \left(\Abf_{22}-\Abf_{21}\Abf_{11}^{-1}\Abf_{12}\right)^{-1}
         \end{array}
\right]\nn\\
&&=\left[\begin{array}{cc}
        \Abf_{11}^{-1} &\quad \0bf \\
        \0bf &\quad \0bf
      \end{array}
\right]+\left[\begin{array}{c}
                \Abf_{11}^{-1}\Abf_{12} \\
                -\Ibf
              \end{array}
\right]\left(\Abf_{22}-\Abf_{21}\Abf_{11}^{-1}\Abf_{12}\right)^{-1}\left[\begin{array}{cc}
                \Abf_{21}\Abf_{11}^{-1} &\quad -\Ibf
              \end{array}
\right].\nn
\eqa
where (a) is by the block matrix inversion lemma \cite[p. 18]{Horn&Johnson:book}, and (b) is by the Woodbury matrix identity \cite[p. 19]{Horn&Johnson:book}:
\bqa
\left(\Cbf+\Ubf\Bbf\Vbf\right)^{-1}=\Cbf^{-1}-\Cbf^{-1}\Ubf\left(\Bbf^{-1}+\Vbf\Cbf^{-1}\Ubf\right)^{-1}\Vbf\Cbf^{-1}.
\label{eq:woodbury}
\eqa
\epf

The noisy-interference sum-rate capacity of a MIMO IC is obtained in the following theorem:
\begin{theorem}
For the MIMO IC defined in (\ref{eq:model}) and $P_i>0, i=1,2$, if the optimal solution of problem (\ref{eq:lowerMIMO}) has $\tr\left(\Sbf_i^*\right)>0$, and there exist matrices $\Abf_i$ and $\Sigmabf_i$ that satisfy (\ref{eq:Ei})-(\ref{eq:sigma2}) and
\bqa
&&\Sbf_1^*\Fbf_1^T=\Sbf_1^*\Hbf_1^T\left(\Ibf+\Fbf_2\Sbf_2^*\Fbf_2^T\right)^{-1}\Abf_1
\label{eq:Markov1}\\
&&\Sbf_2^*\Fbf_2^T=\Sbf_2^*\Hbf_2^T\left(\Ibf+\Fbf_1\Sbf_1^*\Fbf_1^T\right)^{-1}\Abf_2
\label{eq:Markov2}\\
&&\Wbf_1\succeq \Obf_1
\label{eq:W1O1}\\
&&\Wbf_2\succeq \Obf_2
\label{eq:W2O2}
\eqa
where
\bqa
\Wbf_1&{}={}&\Gbf_1-\frac{\tr\left(\Sbf_1^*\Gbf_1\right)}{P_1}\Ibf
\label{eq:W1}\\
\Wbf_2&{}={}&\Gbf_2-\frac{\tr\left(\Sbf_2^*\Gbf_2\right)}{P_2}\Ibf
\label{eq:W2}\\
\Obf_1&{}={}&\frac{1}{2}\left[\Abf_1^T\left(\Ibf+\Fbf_2\Sbf_2^*\Fbf_2^T\right)^{-1}\Hbf_1-\Fbf_1\right]^T\left[\Sigmabf_1-\Abf_1^T\left(\Ibf+\Fbf_2\Sbf_2^*\Fbf_2^T\right)^{-1}\Abf_1\right]^{-1}\nn\\
  &&\cdot\left[\Abf_1^T\left(\Ibf+\Fbf_2\Sbf_2^*\Fbf_2^T\right)^{-1}\Hbf_1-\Fbf_1\right]
\label{eq:Obf1}\\
\Obf_2&{}={}&\frac{1}{2}\left[\Abf_2^T\left(\Ibf+\Fbf_1\Sbf_1^*\Fbf_1^T\right)^{-1}\Hbf_2-\Fbf_2\right]^T\left[\Sigmabf_2-\Abf_2^T\left(\Ibf+\Fbf_1\Sbf_1^*\Fbf_1^T\right)^{-1}\Abf_2\right]^{-1}\nn\\
  &&\cdot\left[\Abf_2^T\left(\Ibf+\Fbf_1\Sbf_1^*\Fbf_1^T\right)^{-1}\Hbf_2-\Fbf_2\right]
\label{eq:Obf2}
\eqa
and $\Gbf_1$ and $\Gbf_2$ are defined in (\ref{eq:G1}) and (\ref{eq:G2}), respectively, then the sum-rate capacity is the maximum in problem (\ref{eq:lowerMIMO}) and is achieved by Gaussian input $\xp_i^*\sim\Nmat\left(0,\Sbf_i^*\right)$ and treating interference as noise.
\label{theorem:MIMO}
\end{theorem}
\bpf
It suffices to show that under conditions (\ref{eq:Ei})-(\ref{eq:sigma2}) and (\ref{eq:Markov1})-(\ref{eq:W2O2}), the upper bound on the sum-rate capacity, i.e., the maximum in problem (\ref{eq:upper}) for the given $\Abf_i$ and $\Sigmabf_i$, is the same as the maximum in problem (\ref{eq:lowerMIMO}); and the maximum in (\ref{eq:upper}) is also achieved by $\Sbf_i^*$.

The proof has two stages. In stage one, we rewrite the objective function of problem (\ref{eq:upper}) and show that this objective function, by choosing $\Sbf_i=\Sbf_i^*$, equals the maximum achieved in problem (\ref{eq:lowerMIMO}). In stage two, we compare the KKT conditions of problems (\ref{eq:lowerMIMO}) and (\ref{eq:upper}), and show that if the conditions in this theorem are all satisfied, then problem (\ref{eq:upper}) is solved by the same $\Sbf_i^*$ that maximizes (\ref{eq:lowerMIMO}).

Define
\bqa
R_{1u}\left(\Sbf_1,\Sbf_2\right)&{}={}&\frac{1}{2}\log\left|\Ibf+\left[\begin{array}{c}
                                                    \Hbf_1 \\
                                                    \Fbf_1
                                                  \end{array}
\right]\Sbf_1\left[\begin{array}{c}
                                                    \Hbf_1 \\
                                                    \Fbf_1
                                                  \end{array}
\right]^T\left(\Ebf_1+\left[\begin{array}{c}
                            \Fbf_2 \\
                            \0bf
                          \end{array}
\right]\Sbf_2\left[\begin{array}{c}
                            \Fbf_2 \\
                            \0bf
                          \end{array}
\right]^T\right)^{-1}\right|\\
R_{2u}\left(\Sbf_1,\Sbf_2\right)&{}={}&\frac{1}{2}\log\left|\Ibf+\left[\begin{array}{c}
                                                    \Hbf_2 \\
                                                    \Fbf_2
                                                  \end{array}
\right]\Sbf_2\left[\begin{array}{c}
                                                    \Hbf_2 \\
                                                    \Fbf_2
                                                  \end{array}
\right]^T\left(\Ebf_2+\left[\begin{array}{c}
                            \Fbf_1 \\
                            \0bf
                          \end{array}
\right]\Sbf_1\left[\begin{array}{c}
                            \Fbf_1 \\
                            \0bf
                          \end{array}
\right]^T\right)^{-1}\right|.
\eqa

Before proceeding, we first show the following equality since it will be used repeatedly in the sequel:
\bqa
&&\left[\begin{array}{c}
                                                    \Hbf_i \\
                                                    \Fbf_i
                                                  \end{array}
\right]^T\left(\Ebf_i+\left[\begin{array}{c}
                            \Fbf_j \\
                            \0bf
                          \end{array}
\right]\Sbf_j\left[\begin{array}{c}
                            \Fbf_j \\
                            \0bf
                          \end{array}
\right]^T\right)^{-1}\left[\begin{array}{c}
                                                    \Hbf_i \\
                                                    \Fbf_i
                                                  \end{array}
\right]\nn\\
&&=\left[\begin{array}{c}
                                                    \Hbf_i \\
                                                    \Fbf_i
                                                  \end{array}
\right]^T\left[\begin{array}{cc}
                            \Ibf+\Fbf_j\Sbf_j\Fbf_j^T &\quad \Abf_i \\
                            \Abf_i^T &\quad \Sigmabf_i
                          \end{array}
\right]^{-1}\left[\begin{array}{c}
                                                    \Hbf_i \\
                                                    \Fbf_i
                                                  \end{array}
\right]\nn\\
&&\stackrel{(a)}=\left[\begin{array}{c}
                                                    \Hbf_i \\
                                                    \Fbf_i
                                                  \end{array}
\right]^T\left(\left[\begin{array}{cc}
                            \left(\Ibf+\Fbf_j\Sbf_j\Fbf_j^T\right)^{-1} &\quad \0bf \\
                            \0bf &\quad \0bf
                          \end{array}
\right]+\left[\begin{array}{c}
                                                    \left(\Ibf+\Fbf_j\Sbf_j\Fbf_j^T\right)^{-1}\Abf_1 \\
                                                    -\Ibf
                                                  \end{array}
\right]\right.\nn\\
  &&\hspace{.6in}\left.\cdot\left[\Sigmabf_i-\Abf_i^T\left(\Ibf+\Fbf_j\Sbf_j\Fbf_j^T\right)^{-1}\Abf_i\right]^{-1}\left[\begin{array}{cc}
                                                    \Abf_1^T\left(\Ibf+\Fbf_j\Sbf_j\Fbf_j^T\right)^{-1} &\quad -\Ibf
                                                  \end{array}
\right]\right)^{-1}\left[\begin{array}{c}
                                                    \Hbf_i \\
                                                    \Fbf_i
                                                  \end{array}
\right]\nn\\
&&=\Hbf_i^T\left(\Ibf+\Fbf_j\Sbf_j\Fbf_j^T\right)^{-1}\Hbf_i+\left[\Abf_i^T\left(\Ibf+\Fbf_j\Sbf_j\Fbf_j^T\right)^{-1}\Hbf_i-\Fbf_i\right]^T\nn\\
  &&\hspace{.6in}\cdot\left[\Sigmabf_i-\Abf_i^T\left(\Ibf+\Fbf_j\Sbf_j\Fbf_j^T\right)^{-1}\Abf_i\right]^{-1}\left[\Abf_i^T\left(\Ibf+\Fbf_j\Sbf_j\Fbf_j^T\right)^{-1}\Hbf_i-\Fbf_i\right]\nn\\
&&=\Hbf_i^T\left(\Ibf+\Fbf_j\Sbf_j\Fbf_j^T\right)^{-1}\Hbf_i+2\overline\Obf_i
\label{eq:impEq}
\eqa
where (a) is by Lemma \ref{lemma:myMxInv}, $i,j\in\{1,2\}$, and $i\neq j$, and we define $\overline\Obf_i$ in the same way as in (\ref{eq:Obf1}) and (\ref{eq:Obf2}) by replacing $\Sbf_i^*$ with $\Sbf_i$.

We first show $R_{il}\left(\Sbf_1^*,\Sbf_2^*\right)=R_{iu}\left(\Sbf_1^*,\Sbf_2^*\right)$:
\bqa
&&R_{1u}\left(\Sbf_1,\Sbf_2\right)\nn\\
&&\stackrel{(a)}=\frac{1}{2}\log\left|\Ibf+\Sbf_1\left[\begin{array}{c}
                                                    \Hbf_1 \\
                                                    \Fbf_1
                                                  \end{array}
\right]^T\left(\Ebf_1+\left[\begin{array}{c}
                            \Fbf_2 \\
                            \0bf
                          \end{array}
\right]\Sbf_2\left[\begin{array}{c}
                            \Fbf_2 \\
                            \0bf
                          \end{array}
\right]^T\right)^{-1}\left[\begin{array}{c}
                                                    \Hbf_1 \\
                                                    \Fbf_1
                                                  \end{array}
\right]\right|\nn\\
&&\stackrel{(b)}=\frac{1}{2}\log\left|\Ibf+\Sbf_1\Hbf_1^T\left(\Ibf+\Fbf_2\Sbf_2\Fbf_2^T\right)^{-1}\Hbf_1+2\Sbf_1\overline\Obf_1\right|
\label{eq:R1u}
\eqa
where (a) is by the matrix identity
\bqa
|\Ibf+\Cbf\Dbf|=|\Ibf+\Dbf\Cbf|
\label{eq:detIdentity}
\eqa
and (b) is from (\ref{eq:impEq}). Similarly, we have
\bqa
&&R_{2u}\left(\Sbf_1,\Sbf_2\right)=\frac{1}{2}\log\left|\Ibf+\Sbf_2\Hbf_2^T\left(\Ibf+\Fbf_1\Sbf_1\Fbf_1^T\right)^{-1}\Hbf_2+2\Sbf_2\overline\Obf_2\right|.
\label{eq:R2u}
\eqa

Since (\ref{eq:Markov1}) and (\ref{eq:Markov2}) imply
\bqa
\Sbf_i^*\Obf_i=\0bf
\label{eq:SOi}
\eqa
then we immediately have
\bqa
R_{1u}\left(\Sbf_1^*,\Sbf_2^*\right)&{}={}&\frac{1}{2}\log\left|\Ibf+\Sbf_1^*\Hbf_1^T\left(\Ibf+\Fbf_2\Sbf_2^*\Fbf_2^T\right)^{-1}\Hbf_1\right|\nn\\
&{}={}&\frac{1}{2}\log\left|\Ibf+\Hbf_1\Sbf_1^*\Hbf_1^T\left(\Ibf+\Fbf_2\Sbf_2^*\Fbf_2^T\right)^{-1}\right|\nn\\
&{}={}&R_{1l}\left(\Sbf_1^*,\Sbf_2^*\right)
\label{eq:eqlbounds1}
\eqa
where the second equality is by (\ref{eq:detIdentity}).
Similarly, we have
\bqa
R_{2u}\left(\Sbf_1^*,\Sbf_2^*\right)=R_{2l}\left(\Sbf_1^*,\Sbf_2^*\right).
\label{eq:eqlbounds2}
\eqa

Next, we prove that the maximum in problem (\ref{eq:upper}) is achieved when $\Sbf_i=\Sbf_i^*$. Since by Lemma \ref{lemma:cvxOpt}, problem (\ref{eq:upper}) is a convex optimization problem, it suffices to prove that there exist Lagrangian multipliers $\overline\lambda_i$ and $\overline\Wbf_i$ such that the following KKT conditions are satisfied:
\bqa
&&-\left.\frac{\partial R_{1u}}{\partial\Sbf_1}\right|_{\Sbf_i=\Sbf_i^*}-\left.\frac{\partial R_{2u}}{\partial\Sbf_1}\right|_{\Sbf_i=\Sbf_i^*}+\overline\lambda_1\Ibf-\overline\Wbf_1=0
\label{eq:upperKKTS1}\\
&&-\left.\frac{\partial R_{1u}}{\partial\Sbf_2}\right|_{\Sbf_i=\Sbf_i^*}-\left.\frac{\partial R_{2u}}{\partial\Sbf_2}\right|_{\Sbf_i=\Sbf_i^*}+\overline\lambda_2\Ibf-\overline\Wbf_2=0
\label{eq:upperKKTS2}\\
&&\overline\lambda_i\left\{\begin{array}{ll}
                     > 0 &\quad \textrm{if } \tr(\Sbf_i^*)=P_i \\
                     =0 &\quad \textrm{if } \tr(\Sbf_i^*)<P_i
                   \end{array}
 \right.\quad i=1,2
\label{eq:upperKKTlambda}\\
&&\tr\left(\Sbf_i^*\overline\Wbf_i\right)=0,\quad \overline\Wbf_i\succeq\0bf.
\label{eq:upperKKTpsd}
\eqa
We first compute
\bqa
&&-\left.\frac{\partial R_{1u}}{\partial\Sbf_1}\right|_{\Sbf_i=\Sbf_i^*}\nn\\
&&\stackrel{(a)}=-\frac{1}{2}\left.\frac{\partial}{\partial\Sbf_1}\left(\log\left|\Ibf+\Sbf_1\left[\begin{array}{c}
                                                    \Hbf_1 \\
                                                    \Fbf_1
                                                  \end{array}
\right]^T\left(\Ebf_1+\left[\begin{array}{c}
                            \Fbf_2 \\
                            \0bf
                          \end{array}
\right]\Sbf_2\left[\begin{array}{c}
                            \Fbf_2 \\
                            \0bf
                          \end{array}
\right]^T\right)^{-1}\left[\begin{array}{c}
                                                    \Hbf_1 \\
                                                    \Fbf_1
                                                  \end{array}
\right]\right|\right)\right|_{\Sbf_i=\Sbf_i^*}\nn\\
&&=-\frac{1}{2}\left[\begin{array}{c}
                                                    \Hbf_1 \\
                                                    \Fbf_1
                                                  \end{array}
\right]^T\left(\Ebf_1+\left[\begin{array}{c}
                            \Fbf_2 \\
                            \0bf
                          \end{array}
\right]\Sbf_2^*\left[\begin{array}{c}
                            \Fbf_2 \\
                            \0bf
                          \end{array}
\right]^T\right)^{-1}\left[\begin{array}{c}
                                                    \Hbf_1 \\
                                                    \Fbf_1
                                                  \end{array}
\right]\nn\\
 &&\hspace{.2in}\cdot\left(\Ibf+\Sbf_1^*\left[\begin{array}{c}
                                                    \Hbf_1 \\
                                                    \Fbf_1
                                                  \end{array}
\right]^T\left(\Ebf_1+\left[\begin{array}{c}
                            \Fbf_2 \\
                            \0bf
                          \end{array}
\right]\Sbf_2^*\left[\begin{array}{c}
                            \Fbf_2 \\
                            \0bf
                          \end{array}
\right]^T\right)^{-1}\left[\begin{array}{c}
                                                    \Hbf_1 \\
                                                    \Fbf_1
                                                  \end{array}
\right]\right)^{-1}\nn\\
&&\stackrel{(b)}=-\frac{1}{2}\left[\begin{array}{c}
                                                    \Hbf_1 \\
                                                    \Fbf_1
                                                  \end{array}
\right]^T\left[\begin{array}{cc}
                            \Ibf+\Fbf_2\Sbf_2^*\Fbf_2^T &\quad \Abf_1 \\
                            \Abf_1^T &\quad \Sigmabf_1
                          \end{array}
\right]^{-1}\left[\begin{array}{c}
                                                    \Hbf_1 \\
                                                    \Fbf_1
                                                  \end{array}
\right]\left(\Ibf+\Sbf_1^*\left(\Hbf_1^T\left(\Ibf+\Fbf_2\Sbf_2^*\Fbf_2^T\right)^{-1}\Hbf_1+2\Obf_1\right)\right)\nn\\
&&\stackrel{(c)}=-\frac{1}{2}\left[\begin{array}{c}
                                                    \Hbf_1 \\
                                                    \Fbf_1
                                                  \end{array}
\right]^T\left[\begin{array}{cc}
                            \Ibf+\Fbf_2\Sbf_2^*\Fbf_2^T &\quad \Abf_1 \\
                            \Abf_1^T &\quad \Sigmabf_1
                          \end{array}
\right]^{-1}\left[\begin{array}{c}
                                                    \Hbf_1 \\
                                                    \Fbf_1
                                                  \end{array}
\right]\left(\Ibf+\Sbf_1^*\Hbf_1^T\left(\Ibf+\Fbf_2\Sbf_2^*\Fbf_2^T\right)^{-1}\Hbf_1\right)^{-1}\nn\\
&&\stackrel{(d)}=-\frac{1}{2}\left(\Hbf_1^T\left(\Ibf+\Fbf_2\Sbf_2^*\Fbf_2^T\right)^{-1}\Hbf_1+2\Obf_1\right)\left(\Ibf+\Sbf_1^*\Hbf_1^T\left(\Ibf+\Fbf_2\Sbf_2^*\Fbf_2^T\right)^{-1}\Hbf_1\right)^{-1}\nn\\
&&=-\frac{1}{2}\Hbf_1^T\left(\Ibf+\Fbf_2\Sbf_2^*\Fbf_2^T\right)^{-1}\Hbf_1\left(\Ibf+\Sbf_1^*\Hbf_1^T\left(\Ibf+\Fbf_2\Sbf_2^*\Fbf_2^T\right)^{-1}\Hbf_1\right)^{-1}\nn\\
  &&\hspace{.6in}-\Obf_1\left(\Ibf+\Sbf_1^*\Hbf_1^T\left(\Ibf+\Fbf_2\Sbf_2^*\Fbf_2^T\right)^{-1}\Hbf_1\right)^{-1}\nn\\
&&\stackrel{(e)}=-\frac{1}{2}\Hbf_1^T\left(\Ibf+\Fbf_2\Sbf_2^*\Fbf_2^T\right)^{-1}\Hbf_1\left(\Ibf+\Sbf_1^*\Hbf_1^T\left(\Ibf+\Fbf_2\Sbf_2^*\Fbf_2^T\right)^{-1}\Hbf_1\right)^{-1}\nn\\
  &&\hspace{.6in}-\Obf_1\left(\Ibf-\Sbf_1^*\Hbf_1^T\left(\Ibf+\Fbf_2\Sbf_2^*\Fbf_2^T+\Hbf_1\Sbf_1^*\Hbf_1^T\right)^{-1}\Hbf_1\right)\nn\\
&&\stackrel{(f)}=-\frac{1}{2}\Hbf_1^T\left(\Ibf+\Fbf_2\Sbf_2^*\Fbf_2^T\right)^{-1}\Hbf_1\left(\Ibf+\Sbf_1^*\Hbf_1^T\left(\Ibf+\Fbf_2\Sbf_2^*\Fbf_2^T\right)^{-1}\Hbf_1\right)^{-1}-\Obf_1\nn\\
&&\stackrel{(g)}=-\frac{1}{2}\Hbf_1^T\left(\Ibf+\Fbf_2\Sbf_2^*\Fbf_2^T\right)^{-1}\left(\Ibf+\Hbf_1\Sbf_1^*\Hbf_1^T\left(\Ibf+\Fbf_2\Sbf_2^*\Fbf_2^T\right)^{-1}\right)^{-1}\Hbf_1-\Obf_1\nn\\
&&=-\frac{1}{2}\Hbf_1^T\left(\Ibf+\Hbf_1\Sbf_1^*\Hbf_1^T+\Fbf_2\Sbf_2^*\Fbf_2^T\right)^{-1}\Hbf_1-\Obf_1\nn\\
&&=-\frac{1}{2}\left.\frac{\partial}{\partial\Sbf_1}\left[\log\left(\Ibf+\Hbf_1\Sbf_1\Hbf_1^T+\Fbf_2\Sbf_2\Fbf_2^T\right)-\log\left(\Ibf+\Fbf_2\Sbf_2\Fbf_2^T\right)\right]\right|_{\Sbf_i=\Sbf_i^*}-\Obf_1\nn\\
&&=-\left.\frac{\partial R_{1l}}{\partial\Sbf_1}\right|_{\Sbf_i=\Sbf_i^*}-\Obf_1
\eqa
where (a) is by the matrix identity (\ref{eq:detIdentity}), (b) and (d) are both by (\ref{eq:impEq}), (c) and (f) are both by (\ref{eq:SOi}), (e) is by the Woodbury matrix identity (\ref{eq:woodbury}),  and (g) is by the matrix identity \cite[p. 151]{Searle:book}:
\bqa
\Cbf\left(\Ibf+\Dbf\Cbf\right)^{-1}=\left(\Ibf+\Cbf\Dbf\right)^{-1}\Cbf.
\label{eq:Searle1}
\eqa
Then we compute
\bqa
&&-\left.\frac{\partial R_{1u}}{\partial\Sbf_2}\right|_{\Sbf_i=\Sbf_i^*}\nn\\
&&=-\frac{1}{2}\left.\frac{\partial}{\partial\Sbf_2}\left(\log\left|\Ebf_1+\left[\begin{array}{c}
                                                    \Hbf_1 \\
                                                    \Fbf_1
                                                  \end{array}
\right]\Sbf_1\left[\begin{array}{c}
                                                    \Hbf_1 \\
                                                    \Fbf_1
                                                  \end{array}
\right]^T+\left[\begin{array}{c}
                            \Fbf_2 \\
                            \0bf
                          \end{array}
\right]\Sbf_2\left[\begin{array}{c}
                            \Fbf_2 \\
                            \0bf
                          \end{array}
\right]^T\right|-\log\left|\Ebf_1+\left[\begin{array}{c}
                            \Fbf_2 \\
                            \0bf
                          \end{array}
\right]\Sbf_2\left[\begin{array}{c}
                            \Fbf_2 \\
                            \0bf
                          \end{array}
\right]^T\right|\right)\right|_{\Sbf_i=\Sbf_i^*}\nn\\
&&=-\frac{1}{2}\left[\begin{array}{c}
                            \Fbf_2 \\
                            \0bf
                          \end{array}
\right]^T\left(\Ebf_1+\left[\begin{array}{c}
                                                    \Hbf_1 \\
                                                    \Fbf_1
                                                  \end{array}
\right]\Sbf_1^*\left[\begin{array}{c}
                                                    \Hbf_1 \\
                                                    \Fbf_1
                                                  \end{array}
\right]^T+\left[\begin{array}{c}
                            \Fbf_2 \\
                            \0bf
                          \end{array}
\right]\Sbf_2^*\left[\begin{array}{c}
                            \Fbf_2 \\
                            \0bf
                          \end{array}
\right]^T\right)^{-1}\left[\begin{array}{c}
                            \Fbf_2 \\
                            \0bf
                          \end{array}
\right]\nn\\
&&\hspace{.6in}+\frac{1}{2}\left[\begin{array}{c}
                            \Fbf_2 \\
                            \0bf
                          \end{array}
\right]^T\left(\Ebf_1+\left[\begin{array}{c}
                            \Fbf_2 \\
                            \0bf
                          \end{array}
\right]\Sbf_2^*\left[\begin{array}{c}
                            \Fbf_2 \\
                            \0bf
                          \end{array}
\right]^T\right)^{-1}\left[\begin{array}{c}
                            \Fbf_2 \\
                            \0bf
                          \end{array}
\right]\nn\\
&&\stackrel{(a)}=\frac{1}{2}\left[\begin{array}{c}
                            \Fbf_2 \\
                            \0bf
                          \end{array}
\right]^T\left[\begin{array}{cc}
                                                    \Ibf+\Hbf_1\Sbf_1^*\Hbf_1^T+\Fbf_2\Sbf_2^*\Fbf_2^T &\quad \Hbf_1\Sbf_1^*\Fbf_1^T+\Abf_1 \\
                                                    \Fbf_1\Sbf_1^*\Hbf_1^T+\Abf_1^T &\quad \Fbf_1\Sbf_1^*\Fbf_1^T+\Sigmabf_1
                                                  \end{array}
\right]^{-1}\left[\begin{array}{c}
                                                    \Hbf_1 \\
                                                    \Fbf_1
                                                  \end{array}
\right]\Sbf_1^*\left[\begin{array}{c}
                                                    \Hbf_1 \\
                                                    \Fbf_1
                                                  \end{array}
\right]^T\nn\\
&&\hspace{.6in}\cdot\left[\begin{array}{cc}
                                                    \Ibf+\Fbf_2\Sbf_2^*\Fbf_2^T &\quad \Abf_1 \\
                                                    \Abf_1^T &\quad \Sigmabf_1
                                                  \end{array}
\right]^{-1}\left[\begin{array}{c}
                            \Fbf_2 \\
                            \0bf
                          \end{array}
\right]\nn\\
&&\stackrel{(b)}=\frac{1}{2}\left[\begin{array}{c}
                            \Fbf_2 \\
                            \0bf
                          \end{array}
\right]^T\left[\begin{array}{cc}
                                                    \Ibf+\Hbf_1\Sbf_1^*\Hbf_1^T+\Fbf_2\Sbf_2^*\Fbf_2^T &\quad \Hbf_1\Sbf_1^*\Fbf_1^T+\Abf_1 \\
                                                    \Fbf_1\Sbf_1^*\Hbf_1^T+\Abf_1^T &\quad \Fbf_1\Sbf_1^*\Fbf_1^T+\Sigmabf_1
                                                  \end{array}
\right]^{-1}\left[\begin{array}{c}
                                                    \Hbf_1 \\
                                                    \Fbf_1
                                                  \end{array}
\right]\Sbf_1^*\left[\begin{array}{c}
                                                    \Hbf_1 \\
                                                    \Fbf_1
                                                  \end{array}
\right]^T\nn\\
  &&\hspace{.6in}\left(\left[\begin{array}{cc}
                               \left(\Ibf+\Fbf_2\Sbf_2^*\Fbf_2^T\right)^{-1} &\quad \0bf \\
                               \0bf &\quad \0bf
                             \end{array}
  \right]+\left[\begin{array}{c}
                               \left(\Ibf+\Fbf_2\Sbf_2^*\Fbf_2^T\right)^{-1}\Abf_1 \\
                               -\Ibf
                             \end{array}
  \right]\left(\Sigmabf_1-\Abf_1^T\left(\Ibf+\Fbf_2\Sbf_2^*\Fbf_2^T\right)^{-1}\Abf_1\right)^{-1}\right.\nn\\
  &&\hspace{.6in}\left.\left[\begin{array}{cc}
                               \Abf_1^T\left(\Ibf+\Fbf_2\Sbf_2^*\Fbf_2^T\right)^{-1} &\quad -\Ibf
                             \end{array}
  \right]\right)\left[\begin{array}{c}
                            \Fbf_2 \\
                            \0bf
                          \end{array}
\right]\nn\\
&&\stackrel{(c)}=\frac{1}{2}\left[\begin{array}{c}
                            \Fbf_2 \\
                            \0bf
                          \end{array}
\right]^T\left[\begin{array}{cc}
                                                    \Ibf+\Hbf_1\Sbf_1^*\Hbf_1^T+\Fbf_2\Sbf_2^*\Fbf_2^T &\quad \Hbf_1\Sbf_1^*\Fbf_1^T+\Abf_1 \\
                                                    \Fbf_1\Sbf_1^*\Hbf_1^T+\Abf_1^T &\quad \Fbf_1\Sbf_1^*\Fbf_1^T+\Sigmabf_1
                                                  \end{array}
\right]^{-1}\left[\begin{array}{c}
                                                    \Hbf_1 \\
                                                    \Fbf_1
                                                  \end{array}
\right]\Sbf_1^*\Hbf_1^T\left(\Ibf+\Fbf_2\Sbf_2^*\Fbf_2^T\right)^{-1}\Fbf_2\nn\\
&&\stackrel{(d)}=\frac{1}{2}\left[\begin{array}{c}
                            \Fbf_2 \\
                            \0bf
                          \end{array}
\right]^T\left(\left[\begin{array}{cc}
                               \left(\Ibf+\Hbf_1\Sbf_1^*\Hbf_1^T+\Fbf_2\Sbf_2^*\Fbf_2^T\right)^{-1} &\quad \0bf \\
                               \0bf &\quad \0bf
                             \end{array}
  \right]+\left[\begin{array}{c}
                               \left(\Ibf+\Hbf_1\Sbf_1^*\Hbf_1^T+\Fbf_2\Sbf_2^*\Fbf_2^T\right)^{-1}\left(\Hbf_1\Sbf_1^*\Fbf_1^T+\Abf_1\right) \\
                               -\Ibf
                             \end{array}
  \right]\right.\nn\\
  &&\hspace{.6in}\left.\left(\Fbf_1\Sbf_1^*\Fbf_1^T+\Sigmabf_1-\left(\Fbf_1\Sbf_1^*\Hbf_1^T+\Abf_1^T\right)\left(\Ibf+\Hbf_1\Sbf_1^*\Hbf_1^T+\Fbf_2\Sbf_2^*\Fbf_2^T\right)^{-1}\left(\Fbf_1\Sbf_1^*\Hbf_1^T+\Abf_1^T\right)^T\right)^{-1}\right.\nn\\
  &&\hspace{.6in}\left.\left[\begin{array}{cc}
                               \left(\Fbf_1\Sbf_1^*\Hbf_1^T+\Abf_1^T\right)\left(\Ibf+\Hbf_1\Sbf_1^*\Hbf_1^T+\Fbf_2\Sbf_2^*\Fbf_2^T\right)^{-1} &\quad-\Ibf
                             \end{array}
  \right]\right)\left[\begin{array}{c}
                                                    \Hbf_1 \\
                                                    \Fbf_1
                                                  \end{array}
\right]\Sbf_1^*\Hbf_1^T\left(\Ibf+\Fbf_2\Sbf_2^*\Fbf_2^T\right)^{-1}\Fbf_2\nn\\
&&=\frac{1}{2}\Fbf_2^T\left(\Ibf+\Hbf_1\Sbf_1^*\Hbf_1^T+\Fbf_2\Sbf_2^*\Fbf_2^T\right)^{-1}\Hbf_1\Sbf_1^*\Hbf_1^T\left(\Ibf+\Fbf_2\Sbf_2^*\Fbf_2^T\right)^{-1}\Fbf_2\nn\\
&&\hspace{.6in}+\frac{1}{2}\left[\begin{array}{c}
                            \Fbf_2 \\
                            \0bf
                          \end{array}
\right]^T\left[\begin{array}{c}
                               \left(\Ibf+\Hbf_1\Sbf_1^*\Hbf_1^T+\Fbf_2\Sbf_2^*\Fbf_2^T\right)^{-1}\left(\Hbf_1\Sbf_1^*\Fbf_1^T+\Abf_1\right) \\
                               -\Ibf
                             \end{array}
  \right]\nn\\
&&\hspace{.6in}\cdot\left(\Fbf_1\Sbf_1^*\Fbf_1^T+\Sigmabf_1-\left(\Fbf_1\Sbf_1^*\Hbf_1^T+\Abf_1^T\right)\left(\Ibf+\Hbf_1\Sbf_1^*\Hbf_1^T+\Fbf_2\Sbf_2^*\Fbf_2^T\right)^{-1}\left(\Fbf_1\Sbf_1^*\Hbf_1^T+\Abf_1^T\right)^T\right)^{-1}\nn\\
&&\hspace{.6in}\cdot\left(\left(\Fbf_1\Sbf_1^*\Hbf_1^T+\Abf_1^T\right)\left(\Ibf+\Hbf_1\Sbf_1^*\Hbf_1^T+\Fbf_2\Sbf_2^*\Fbf_2^T\right)^{-1}\Hbf_1-\Fbf_1\right)\Sbf_1^*\Hbf_1^T\left(\Ibf+\Fbf_2\Sbf_2^*\Fbf_2^T\right)^{-1}\Fbf_2\nn\\
&&\stackrel{(e)}=\frac{1}{2}\Fbf_2^T\left(\Ibf+\Hbf_1\Sbf_1^*\Hbf_1^T+\Fbf_2\Sbf_2^*\Fbf_2^T\right)^{-1}\Hbf_1\Sbf_1^*\Hbf_1^T\left(\Ibf+\Fbf_2\Sbf_2^*\Fbf_2^T\right)^{-1}\Fbf_2\nn\\
&&\stackrel{(f)}=-\frac{1}{2}\Fbf_2^T\left(\Ibf+\Hbf_1\Sbf_1^*\Hbf_1^T+\Fbf_2\Sbf_2^*\Fbf_2^T\right)^{-1}\Fbf_2+\frac{1}{2}\Fbf_2^T\left(\Ibf+\Fbf_2\Sbf_2^*\Fbf_2^T\right)^{-1}\Fbf_2\nn\\
&&=-\frac{1}{2}\left.\frac{\partial}{\partial\Sbf_2}\left[\log\left(\Ibf+\Hbf_1\Sbf_1\Hbf_1^T+\Fbf_2\Sbf_2\Fbf_2^T\right)-\log\left(\Ibf+\Fbf_2\Sbf_2\Fbf_2^T\right)\right]\right|_{\Sbf_i=\Sbf_i}\nn\\
&&=-\left.\frac{\partial R_{1l}}{\partial\Sbf_2}\right|_{\Sbf_i=\Sbf_i^*}
\eqa
where both (a) and (f) are from the matrix identity
\bqa
\Cbf^{-1}-\Dbf^{-1}=\Cbf^{-1}\left(\Dbf-\Cbf\right)\Dbf^{-1},
\eqa
equality (b) and (d) are both from Lemma \ref{lemma:myMxInv}, (c) is directly from (\ref{eq:Markov1}), and (e) is also from (\ref{eq:Markov1}) which implies
\bqa
&&\left(\left(\Fbf_1\Sbf_1^*\Hbf_1^T+\Abf_1^T\right)\left(\Ibf+\Hbf_1\Sbf_1^*\Hbf_1^T+\Fbf_2\Sbf_2^*\Fbf_2^T\right)^{-1}\Hbf_1-\Fbf_1\right)\Sbf_1^*\nn\\
&&=\left(\Abf_1^T\left(\Ibf+\Fbf_2\Sbf_2^*\Fbf_2^T\right)^{-1}\Hbf_1\Sbf_1\Hbf_1^T+\Abf_1^T\right)\left(\Ibf+\Hbf_1\Sbf_1^*\Hbf_1^T+\Fbf_2\Sbf_2^*\Fbf_2^T\right)^{-1}\Hbf_1\Sbf_1^*-\Fbf_1\Sbf_1^*\nn\\
&&=\Abf_1^T\left(\Ibf+\Fbf_2\Sbf_2^*\Fbf_2^T\right)^{-1}\Hbf_1\Sbf_1^*-\Fbf_1\Sbf_1^*\nn\\
&&=\0bf.
\eqa
Similarly, we have
\bqa
-\left.\frac{\partial R_{2u}}{\partial\Sbf_1}\right|_{\Sbf_i=\Sbf_i^*}&{}={}&-\left.\frac{\partial R_{2l}}{\partial\Sbf_1}\right|_{\Sbf_i=\Sbf_i^*}\\
-\left.\frac{\partial R_{2u}}{\partial\Sbf_2}\right|_{\Sbf_i=\Sbf_i^*}&{}={}&-\left.\frac{\partial R_{2l}}{\partial\Sbf_2}\right|_{\Sbf_i=\Sbf_i^*}-\Obf_2.
\eqa
By (\ref{eq:lowerKKTS1}) and (\ref{eq:lowerIneq}), we have
\bqa
\Sbf_i^*\Gbf_i+\lambda_i\Sbf_i^*=0.
\eqa
Thus, by (\ref{eq:lowerEq}) we have
\bqa
\lambda_i=-\frac{\tr\left(\Sbf_i^*\Gbf_i\right)}{P_i},
\eqa
and hence from (\ref{eq:lowerKKTS1}) and (\ref{eq:lowerKKTS2}) we have
\bqa
\Wbf_i=\Gbf_i-\frac{\tr\left(\Sbf_i^*\Gbf_i\right)}{P_i}\Ibf
\eqa
i.e., the $\Wbf_i$'s defined in (\ref{eq:W1}) and $(\ref{eq:W2})$ are the Lagrangian multipliers in (\ref{eq:lowerKKTS1}) and (\ref{eq:lowerKKTS2}).

Then, we choose
\bqa
\overline\lambda_i&{}={}&\lambda_i\\
\overline\Wbf_i&{}={}&\Wbf_i-\Obf_i
\eqa
such that
\bqa
&&-\left.\frac{\partial R_{1u}}{\partial\Sbf_1}\right|_{\Sbf_i=\Sbf_i^*}-\left.\frac{\partial R_{2u}}{\partial\Sbf_1}\right|_{\Sbf_i=\Sbf_i^*}+\overline\lambda_1\Ibf-\overline\Wbf_1\nn\\
&&=-\left.\frac{\partial R_{1l}}{\partial\Sbf_1}\right|_{\Sbf_i=\Sbf_i^*}-\Obf_1-\left.\frac{\partial R_{2l}}{\partial\Sbf_1}\right|_{\Sbf_i=\Sbf_i^*}+\lambda_1\Ibf-\left(\Wbf_1-\Obf_1\right)\nn\\
&&=\0bf
\eqa
where the last equality is from (\ref{eq:lowerKKTS1}). Therefore, condition (\ref{eq:upperKKTS1}) is satisfied. Similarly, condition (\ref{eq:upperKKTS2}) is also satisfied. Condition (\ref{eq:upperKKTlambda}) is satisfied because of (\ref{eq:lowerEq}), and condition (\ref{eq:upperKKTpsd}) is satisfied by the assumptions (\ref{eq:W1O1}) and (\ref{eq:W2O2}) and conditions (\ref{eq:Markov1}) and (\ref{eq:Markov2}) which imply
\bqa
\Sbf_i^*\overline\Wbf_i=\Sbf_i^*(\Wbf_i-\Obf_i)=-\Sbf_i^*\Obf_i=\0bf
\eqa
where in the second equality, we use the fact that $\Sbf_i^*\Wbf_i=\0bf$ when $\tr\left(\Sbf_i^*\Wbf_i\right)=0$ and $\Sbf_i^*\succeq\0bf$ and $\Wbf_i\succeq\0bf$. Therefore, there exist Lagrangian multipliers such that $\Sbf_i^*$ satisfies the KKT conditions for problem (\ref{eq:upper}). Since problem (\ref{eq:upper}) is a convex optimization problem, $\Sbf_i^*$ achieves the maximum in problem (\ref{eq:upper}). By (\ref{eq:eqlbounds1}) and (\ref{eq:eqlbounds2}), we conclude that the maximum in (\ref{eq:lowerMIMO}) is the sum-rate capacity of the MIMO IC.
\epf
\begin{remark}
On comparing the upper bound function $R_{ui}$ in (\ref{eq:R1u}) and (\ref{eq:R2u}) with the lower bound function in (\ref{eq:R1l}) and (\ref{eq:R2l}), respectively, we note that there is an extra term $2\Sbf_i\overline\Obf_i$ in the logarithm function. It is obvious that $\overline\Obf_i\succeq\0bf$ under conditions (\ref{eq:sigma1}) and (\ref{eq:sigma2}). Although $2\Sbf_i\overline\Obf_i$ may not necessary be a semi-positive definite matrix, this extra term still increases the rate upon $R_{il}$, e.g.,
\bqn
R_{1u}&{}={}&\frac{1}{2}\log\left|\Ibf+\Sbf_1\Hbf_1^T\left(\Ibf+\Fbf_2\Sbf_2\Fbf_2^T\right)^{-1}\Hbf_1+2\Sbf_1\overline\Obf_1\right|\\
&{}={}&\frac{1}{2}\log\left|\Ibf+\Sbf_1^{\frac{1}{2}}\left(\Hbf_1^T\left(\Ibf+\Fbf_2\Sbf_2\Fbf_2^T\right)^{-1}\Hbf_1+2\overline\Obf_1\right)\Sbf_1^{\frac{1}{2}}\right|\\
&{}\geq{}&\frac{1}{2}\log\left|\Ibf+\Sbf_1^{\frac{1}{2}}\Hbf_1^T\left(\Ibf+\Fbf_2\Sbf_2\Fbf_2^T\right)^{-1}\Hbf_1\Sbf_1^{\frac{1}{2}}\right|\\
&{}={}&\frac{1}{2}\log\left|\Ibf+\Hbf_1\Sbf_1\Hbf_1^T\left(\Ibf+\Fbf_2\Sbf_2\Fbf_2^T\right)^{-1}\right|\\
&{}={}&R_{1l}.
\eqn
Conditions (\ref{eq:Markov1}) and (\ref{eq:Markov2}) are sufficient conditions for (\ref{eq:SOi}) to hold, which makes the lower and upper bounds converge. This extra term $2\Sbf_i\overline\Obf_i$ is also considered in the scaler Gaussian IC \cite[p. 696]{Shang-etal:09IT} and the parallel Gaussian IC \cite[eq. (64)]{Shang-etal:11IT_pgic}, in which we have $\Obf_i=\0bf$ for both cases. Furthermore, conditions (\ref{eq:Markov1}) and (\ref{eq:Markov2}) also mean that \cite[Lemma 5]{Shang-etal:10IT_mimo}
\bqn
\xp_{iG}^*\rightarrow\Hbf_i\xp_{iG}^*+\Fbf_j\xp_{jG}^*+\zp_i\rightarrow \Fbf_i\xp_{iG}^*+\np_i\quad i,j\in\{1,2\}, i\neq j
\eqn
form a Markov chain, where $\xp_{iG}^*\sim\Nmat(\0bf,\Sbf_i^*)$.
\end{remark}

\begin{remark}
When all the conditions in Theorem \ref{theorem:MIMO} are satisfied, the optimal input covariance matrix $\Sbf_i^*$ and the corresponding auxiliary matrix $\Ebf_i^*$ in (\ref{eq:Ei}) (obtained by replacing $\Sigmabf_i$ and $\Abf_i$ with $\Sigmabf_i^*$ and $\Abf_i^*$ associated with $\Sbf_i^*$), form a saddle point of the upper bound function defined as
\bqa
R_{su}(\Sbf_i,\Ebf_i)=R_{1u}(\Sbf_i,\Ebf_i)+R_{2u}(\Sbf_i,\Ebf_i)\nn
\eqa
where $R_{iu}\left(\Sbf_i,\Ebf_i\right)$ is defined in (\ref{eq:R1u}) and (\ref{eq:R2u}). We use this expression in this remark to emphasize that $\Ebf_i$ is also a parameter.

To show that this optimal solution is the saddle point, we first have
\bqa
\min_{\Ebf_i}\max_{\tr(\Sbf_i)\leq P_i}R_{su}(\Sbf_i,\Ebf_i)\leq \max_{\tr(\Sbf_i)\leq P_i}R_{su}(\Sbf_i,\Ebf_i^*)=R_{su}(\Sbf_i^*,\Ebf_i^*)\nn
\eqa
where the second equality is by the existence of the Lagrangian multiplier satisfying the KKT conditions, and the convexity of $R_{su}(\Sbf_i,\Ebf_i^*)$ over $\Sbf_i$, which imply that $R_{su}(\Sbf_i,\Ebf_i^*)$ is maximized by $\Sbf_i^*$. On the other hand, we have
\bqa
\max_{\tr(\Sbf_i)\leq P_i}\min_{\Ebf_i}R_{su}(\Sbf_i,\Ebf_i)\geq \min_{\Ebf_i}R_{su}(\Sbf_i^*,\Ebf_i)=R_{su}(\Sbf_i^*,\Ebf_i^*)\nn
\eqa
where the second inequality is by (\ref{eq:SOi}). Since the following is always true
\bqa
\min_{\Ebf_i}\max_{\tr(\Sbf_i)\leq P_i}R_{su}(\Sbf_i,\Ebf_i)\geq\max_{\tr(\Sbf_i)\leq P_i}\min_{\Ebf_i}R_{su}(\Sbf_i,\Ebf_i)\nn
\eqa
we have
\bqa
\min_{\Ebf_i}\max_{\tr(\Sbf_i)\leq P_i}R_{su}(\Sbf_i,\Ebf_i)=\max_{\tr(\Sbf_i)\leq P_i}\min_{\Ebf_i}R_{su}(\Sbf_i,\Ebf_i)=R_{su}(\Sbf_i^*,\Ebf_i^*).\nn
\eqa
By \cite[Proposition 2.6.1 p. 132]{Bertsekas-etal:book}, $(\Sbf_i^*,\Ebf_i^*)$ is the saddle point of $R_{su}\left(\Sbf_i,\Ebf_i\right)$.
\label{remark:saddle}
\end{remark}
\begin{remark}
Denote by $\bar\Sbf_i$ the covariance matrix constraint in \cite[Theorem 6]{Shang-etal:10IT_mimo} and denote by $\bar\Ebf_i$ the corresponding auxiliary matrix consisting of $\bar\Abf_i$ and $\bar\Sigmabf_i$ for this $\bar\Sbf_i$ that satisfy condition (\ref{eq:Ei})-(\ref{eq:sigma2}), (\ref{eq:Markov1}) and (\ref{eq:Markov2}). If all the conditions in \cite[Theorem 6]{Shang-etal:10IT_mimo} are satisfied, i.e., for any $\0bf\preceq\Sbf_i\preceq\bar\Sbf_i$ there exist corresponding $\Abf_i$ and $\Sigmabf_i$ such that (\ref{eq:Ei})-(\ref{eq:sigma2}), (\ref{eq:Markov1}) and (\ref{eq:Markov2}) are satisfied, then $(\bar\Sbf_i,\bar\Ebf_i)$ is also a saddle point of the upper bound function according to the covariance matrix constraint. This can be shown in a similar way as the result in Remark \ref{remark:saddle}. First, we have
\bqa
\max_{\0bf\preceq\Sbf_i\preceq\bar\Sbf_i}\min_{\Ebf_i}R_{su}(\Sbf_i,\Ebf_i)=\max_{\0bf\preceq\Sbf_i\preceq\bar\Sbf_i}R_{su}(\Sbf_i,\Ebf_i(\Sbf_i))=R_{su}(\bar\Sbf_i,\bar\Abf_i)\nn
\eqa
where the first equality is by the assumption of existence of $\Abf_i$ and $\Sigmabf_i$ that satisfy condition (\ref{eq:Ei})-(\ref{eq:sigma2}), (\ref{eq:Markov1}) and (\ref{eq:Markov2}) for each feasible $\Sbf_i$, and we denote such auxiliary matrix $\Ebf_i$ as $\Ebf_i(\Sbf_i)$. The second equality is by the fact that $R_{su}$ is an increasing function of $\Sbf_i$. On the other hand, we have
\bqa
\min_{\Ebf_i}\max_{\Sbf_i\preceq\bar\Sbf_i}R_{su}(\Sbf_i,\Ebf_i)=\min_{\Ebf_i}R_{su}(\bar\Sbf_i,\Ebf_i)=R_{su}(\bar\Sbf_i,\bar\Ebf_i).\nn
\eqa
Therefore, we also have
\bqa
\max_{\Sbf_i\preceq\bar\Sbf_i}\min_{\Ebf_i}R_{su}(\Sbf_i,\Ebf_i)=\min_{\Ebf_i}\max_{\Sbf_i\preceq\bar\Sbf_i}R_{su}(\Sbf_i,\Ebf_i).\nn
\eqa
By \cite[Proposition 2.6.1 p. 132]{Bertsekas-etal:book}, $(\bar\Sbf_i,\bar\Ebf_i)$ is also a saddle point for $R_{su}(\Sbf_i,\Ebf_i)$ according to the covariance matrix constraint. Therefore, \cite[Theorem 6]{Shang-etal:10IT_mimo} parallels Theorem \ref{theorem:MIMO} in the covariance matrix constraint.
\end{remark}
\begin{remark}
Theorem \ref{theorem:MIMO} includes \cite[Theorem 1]{Annapureddy&Veeravalli:09IT_submission} as a special case. In \cite[Theorem 1]{Annapureddy&Veeravalli:09IT_submission}, it was shown that if the $\Sbf_i^*$ is full rank and there exist $\Abf_i$ and $\Sigmabf_i$ satisfying (\ref{eq:Ei})-(\ref{eq:sigma1}), (\ref{eq:Markov1}) and (\ref{eq:Markov2}), then this MIMO IC has noisy interference. In this case, (\ref{eq:Markov1}) and (\ref{eq:Markov2}) imply
\bqn
\Fbf_i^T=\Hbf_i^T\left(\Ibf+\Fbf_j\Sbf_j^*\Fbf_j^T\right)^{-1}\Abf_i,\quad i,j\in\{1,2\},i\neq j
\eqn
and thus
\bqn
\Obf_i=\0bf.
\eqn
 Therefore, (\ref{eq:W1O1}) and (\ref{eq:W2O2}) are both satisfied since $\Wbf_i\succeq\0bf$ has been shown in Lemma \ref{lemma:KKT}.
\end{remark}
\begin{remark}
Theorem \ref{theorem:MIMO} includes the noisy-interference sum-rate capacity result for the parallel IC in \cite{Shang-etal:11IT_pgic} as a special case. The parallel IC is a special MIMO IC with diagonal channel matrices $\Hbf_i=\diag\left[h_{i1},\cdots,h_{it}\right]$ and $\Fbf_i=\diag\left[f_{i1},\cdots,f_{it}\right]$. We define the $i$th subchannel as that consisting of only the $i$th transmit and receive antennas. The lower bound in (\ref{eq:lowerMIMO}) for this channel, by choosing the diagonal input covariance matrix $\Sbf_i$ can be written as
\bqa
\max&&\quad R_{sl}(\Sbf_i)=\sum_{j=1}^tr_j\left(s_{1j},s_{2j}\right)\nn\\
\textrm{subject to}&&\quad \sum_{j=1}^{t}s_{ij}\leq P_i,\quad s_{ij}\geq 0,\quad i=1,2.
\label{eq:pgicLowernouse}
\eqa
where
\bqa
r_j\left(s_{1j},s_{2j}\right)=\frac{1}{2}\log\left(1+\frac{h_{1j}^2s_{1j}}{1+f_{2j}^2s_{2j}}\right)
+\frac{1}{2}\log\left(1+\frac{h_{2j}^2s_{2j}}{1+f_{1j}^2s_{1j}}\right).
\eqa
However, in \cite{Shang-etal:11IT_pgic} the lower bound on the sum-rate capacity is not formulated as above, but as
\bqa
\max&&\quad \sum_{j=1}^tC_j\left(s_{1j},s_{2j}\right)\nn\\
\textrm{subject to}&&\quad \sum_{j=1}^{t}s_{ij}\leq P_i,\quad s_{ij}\geq 0,\quad i=1,2
\label{eq:lowerPGIC}
\eqa
where $s_{ij}$ denotes the power allocated to the $j$th subchannel for user $i$, and $C_j\left(s_{1j},s_{2j}\right)$ denotes the sum-rate capacity of the $j$th subchannel under power constraint $s_{ij}$, i.e., power $s_{ij}$ is allocated to the $j$th transmit antenna of user $i$. The upper bound on the sum-rate capacity is also formulated via optimization problem (\ref{eq:upper}). However, if we choose the auxiliary matrices $\Abf_i$ and $\Sigmabf_i$ as in \cite[eqs.(41) and (42)]{Shang-etal:11IT_pgic}, then the upper bound can be written as
\bqa
\max&&\quad R_{su}(\Sbf_i)=\sum_{j=1}^tf_j\left(s_{1j},s_{2j}\right)\nn\\
\textrm{subject to}&&\quad \tr(\Sbf_i)=\sum_{j=1}^{t}s_{ij}\leq P_i,\quad s_{ij}\geq 0,\quad i=1,2.
\eqa
where $\Sbf_i=\diag[s_{i1},\cdots,s_{i,t_i}]$ and $f_j(\cdot)$ is defined in \cite[eq.(64)]{Shang-etal:11IT_pgic}. The auxiliary matrix $\Ebf_i$ is the same in both upper bounds. Therefore, \cite{Shang-etal:11IT_pgic} uses exactly the same side information as that in Theorem \ref{theorem:MIMO}. Moreover, \cite{Shang-etal:11IT_pgic} shows that the matrices $\Abf_i^*$ and $\Sigmabf_i^*$ are both diagonal matrices (see $\Ebf_i$ in \cite[eqs. (41) and (42)]{Shang-etal:11IT_pgic}). Thus, the upper bound $R_{su}(\Sbf_i)$ is the sum of the upper bound for each subchannel $f_j$.

It has been shown in \cite{Shang-etal:11IT_pgic} that if the power constraint $P_i$ is in the set \cite[eq. (18)]{Shang-etal:11IT_pgic}, then by \cite[Theorem 3]{Shang-etal:11IT_pgic} this parallel IC has noisy interference and the optimal input covariance matrix $\Sbf_i^*=\diag[s_{i1}^*,\cdots,s_{i,t_i}^*]$ has the properties \cite[eqs.(18), (74) and (75)]{Shang-etal:11IT_pgic}
\bqa
&&\left[\begin{array}{c}
        \lambda_1 \\
        \lambda_2
      \end{array}
\right]\in\bigcap_{j=1}^t\partial C_j\left(s_{1j}^*,s_{2j}^*\right)\neq \textrm{empty}
\label{eq:parallelSubDiff}\\
&&\frac{\partial f_{j}}{\partial s_{ij}}\left|_{\tiny\begin{array}{l}
                                                         s_{1j}=s_{1j}^* \\
                                                         s_{2j}=s_{2j}^*
                                                       \end{array}
}\right.=\frac{\partial C_{j}}{\partial s_{ij}}\left|_{\tiny\begin{array}{l}
                                                         s_{1j}=s_{1j}^* \\
                                                         s_{2j}=s_{2j}^*
                                                       \end{array}
}\right.=\frac{\partial r_{j}}{\partial s_{ij}}\left|_{\tiny\begin{array}{l}
                                                         s_{1j}=s_{1j}^* \\
                                                         s_{2j}=s_{2j}^*
                                                       \end{array}
}\right.\quad\textrm{for all }i=1,2,\quad j=1,\cdots,t
\label{eq:parallelDiff}
\eqa
where $\partial C_j\left(s_{1j}^*,s_{2j}^*\right)$ is the subdifferential of $C_j\left(s_{1j},s_{2j}\right)$ at $\left(s_{1j}^*,s_{2j}^*\right)$, and
$[\lambda_1,\lambda_2]^T$ is the common subgradient shared by all the subdifferentials. From the expression of $\partial C_j\left(s_{1j}^*,s_{2j}^*\right)$ in \cite[eq. (100)]{Shang-etal:11IT_pgic}, we have
\bqa
\left[\begin{array}{c}
        \lambda_1 \\
        \lambda_2
      \end{array}
\right]&{}={}&
\left[\begin{array}{c}
        \left.\dfrac{\partial r_j}{\partial s_{1j}}\right|_{s_{ij}=s_{ij}^*} \\
        \left.\dfrac{\partial r_j}{\partial s_{2j}}\right|_{s_{ij}=s_{ij}^*}
      \end{array}
\right]+\left[\begin{array}{c}
        w_{1j} \\
        w_{2j}
      \end{array}
\right],\quad i=1,2,\quad j=1,\cdots, t
\eqa
where $w_{1j}$ and $w_{2j}$ are nonnegative constants. Hence, we have
\bqa
\left.\frac{\partial R_{sl}}{\partial \Sbf_i}\right|_{\Sbf_1=\Sbf_1^*,\Sbf_2=\Sbf_2^*}=\lambda_i\Ibf+\Wbf_i
\eqa
where $\Wbf_i=\diag[w_{i1},\cdots,w_{it}]\succeq\0bf$. By (\ref{eq:parallelDiff}), we have
\bqa
\left.\frac{\partial R_{su}}{\partial\Sbf_i}\right|_{\Sbf_1=\Sbf_1^*,\Sbf_2=\Sbf_2^*}=\left.\frac{\partial R_{sl}}{\partial \Sbf_i}\right|_{\Sbf_1=\Sbf_1^*,\Sbf_2=\Sbf_2^*}=\lambda_i\Ibf+\Wbf_i
\eqa
which implies $\Obf_i=\0bf$. Therefore, if a parallel IC satisfies the noisy-interference condition in \cite{Shang-etal:11IT_pgic}, it also satisfies Theorem \ref{theorem:MIMO}. The lower bound $\max R_{sl}$ and the upper bound $\max R_{su}$ are optimized at the same $\Sbf_i^*$ with the same Lagrangian multipliers. The Lagrangian multipliers $\lambda_i$ associated with the power constraint $\tr(\Sbf_i)\leq P_i$ form the common subgradient of all the individual subchannel capacities $C_j$ (as well as the individual lower bounds $r_j$) and upper bounds $f_j$, i.e., $C_j$ (or $r_j$) and $f_j$ have parallel supporting hyperplanes with the subgradient $[\lambda_1,\lambda_2]^T$ at the optimal power allocation point.

We note that to formulate the lower bound as in (\ref{eq:lowerPGIC}) is important for \cite{Shang-etal:11IT_pgic} since the problem is then a convex optimization problem. Furthermore, condition (\ref{eq:parallelDiff}) directly guarantees the optimality of $s_{ij}^*$ for (\ref{eq:lowerPGIC}), and only through which we show the optimality of $s_{ij}^*$ for (\ref{eq:pgicLowernouse}) \cite{Shang-etal:11IT_pgic}.
\label{remark:pgic}
\end{remark}
\begin{remark}
Theorem \ref{theorem:MIMO} determines the noisy-interference sum-rate capacity for general MIMO ICs. When the MIMO IC reduces to a MISO or SIMO IC, the conditions in Theorem \ref{theorem:MIMO} can be simplified. We defer these results in Sections \ref{section:MISO} and \ref{section:SIMO}, respectively. In \cite{Annapureddy&Veeravalli:09IT_submission}, noisy-interference sum-rate capacities of symmetric MISO and SIMO ICs are obtained, i.e., ICs with $\Hbf_1=\Hbf_2$, $\Fbf_1=\Fbf_2$, $P_1=P_2$, and where all the $\Hbf_i$ and $\Fbf_i$ are column or row vectors. These two results are both included as special cases of Theorem \ref{theorem:MIMO}. In Sections \ref{section:MISO} and \ref{section:SIMO}, the MISO and SIMO ICs can be symmetric and asymmetric.
\end{remark}
\begin{remark}
Equations (\ref{eq:Markov1}) and (\ref{eq:Markov2}) are special cases of the Sylvester equation \cite{Bartels&Stewart:72ACM}. Once $\Sbf_i^*$ is obtained, the matrix $\Abf_i$  can be obtained by solving the following linear equations:
\bqa
&&\Ibf\otimes\left(\Sbf_1^*\Hbf_1^T\left(\Ibf+\Fbf_2\Sbf_2^*\Fbf_2^T\right)^{-1}\right)\Vec(\Abf_1)=
\Vec\left(\Sbf_1^*\Fbf_1^T\right)\nn\\
&&\Ibf\otimes\left(\Sbf_2^*\Hbf_2^T\left(\Ibf+\Fbf_1\Sbf_1^*\Fbf_1^T\right)^{-1}\right)\Vec(\Abf_2)=
\Vec\left(\Sbf_2^*\Fbf_2^T\right)\nn
\eqa
where  $\otimes$ denotes the Kronecker product of matrices. Therefore, the existence of $\Abf_i$ can be determined by the theory of linear equations.
\label{remark:Sylvester}
\end{remark}
\begin{remark}
In Theorem \ref{theorem:MIMO} and its proof, we need to determine the existence of a positive definite $\Sigmabf_i$. Sometimes the expression for $\Sigmabf_i$ is not important (e.g., the parallel Gaussian IC discussed in Remark \ref{remark:pgic}, and the symmetric SIMO IC discussed later in Remark \ref{remark:sSIMO}). If we choose equality in both (\ref{eq:sigma1}) and (\ref{eq:sigma2}), we obtain two matrix equations which are special cases of a discrete algebraic Ricatti equation \cite{Engwerda-etal:93LA&A}. The existence of a positive definite solution is determined by \cite[Lemma 9]{Shang-etal:10IT_mimo} using \cite{Engwerda-etal:93LA&A}, which requires, for both $i=1$ and $2$:
\bqa
\textrm{radius}\left(\Phibf_i\right)\leq\frac{1}{2}
\label{eq:cond}
\eqa
where
\bqa%
\Phibf_1&{}={}&\left(\Ibf-\Abf_1^T\Abf_1-\Abf_2\Abf_2^T\right)^{-\frac{1}{2}}\Abf_1^T\Abf_2^T
\left(\Ibf-\Abf_1^T\Abf_1-\Abf_2\Abf_2^T\right)^{-\frac{1}{2}}
\label{eq:phi1}\\
\Phibf_2&{}={}&\left(\Ibf-\Abf_1\Abf_1^T-\Abf_2^T\Abf_2\right)^{-\frac{1}{2}}\Abf_2^T\Abf_1^T
\left(\Ibf-\Abf_1\Abf_1^T-\Abf_2^T\Abf_2\right)^{-\frac{1}{2}}.
\label{eq:phi2}
\eqa
Here we present a strengthened result of \cite[Lemma 9]{Shang-etal:10IT_mimo} which requires (\ref{eq:cond}) to be satisfied for only $i=1$ or $i=2$.
\begin{lemma}
For the following matrix equations for $\Sigmabf_1$ and $\Sigmabf_2$:
\bqa
\Sigmabf_1&{}={}&\Ibf-\Abf_2\Sigmabf_2^{-1}\Abf_2^T
\label{eq:RicattiEq1}\\
\Sigmabf_2&{}={}&\Ibf-\Abf_1\Sigmabf_1^{-1}\Abf_1^T
\label{eq:RicattiEq2}
\eqa
if $\textrm{radius}(\Phibf_1)\leq \frac{1}{2}$ or $\textrm{radius}(\Phibf_2)\leq \frac{1}{2}$ where $\Phibf_i$ is defined in (\ref{eq:phi1}) and (\ref{eq:phi2}), then there exist symmetric positive definite solutions for $\Sigmabf_1$ and $\Sigmabf_2$. Moreover, the solutions for both $i=1$ and $2$ satisfy
\bqa
\Sigmabf_i\succ\Abf_i^T\Abf_i
\label{eq:SigmabfAA}
\eqa
or equivalently
\bqa
\Ebf_i=\left[\begin{array}{cc}
        \Ibf &\quad \Abf_i \\
        \Abf_i^T &\quad \Sigmabf_i
      \end{array}
\right]\succ\0bf.
\label{eq:COVfeasible}
\eqa
\end{lemma}
The proof is included in Appendix \ref{appendix:mxEq}.
\label{remark:Ricatti}
\end{remark}

For completeness, we give the noisy-interference condition of MIMO ZIC in the following proposition.
\begin{proposition}
For the MIMO IC defined in (\ref{eq:model}) with $\Fbf_1=\0bf$ and $P_i>0, i=1,2$, if the optimal solution of problem (\ref{eq:lowerMIMO}) has $\tr\left(\Sbf_i^*\right)>0$, and there exist matrices $\Abf_2$ and $\Sigmabf_2$ that satisfy
\bqn
&&\Ibf\succeq\Abf_2\Abf_2^T
\\
&&\Sbf_2^*\Fbf_2^T=\Sbf_2^*\Hbf_2^T\Abf_2
\\
&&\Wbf_2\succeq \Obf_2
\eqn
where
\bqn
\Wbf_2&{}={}&\Gbf_2-\frac{\tr\left(\Sbf_2^*\Gbf_2\right)}{P_2}\Ibf
\\
\Obf_2&{}={}&\frac{1}{2}\left(\Abf_2^T\Hbf_2-\Fbf_2\right)^T\left(\Ibf-\Abf_2^T\Abf_2\right)^{-1}
\left(\Abf_2^T\Hbf_2-\Fbf_2\right)
\eqn
and $\Gbf_2$ are defined in (\ref{eq:G2}), then the sum-rate capacity is the maximum in problem (\ref{eq:lowerMIMO}) and is achieved by Gaussian input $\xp_i^*\sim\Nmat\left(0,\Sbf_i^*\right)$ and treating interference as noise.
\label{prop:MIMOZ}
\end{proposition}
\bpf
The proof is straightforward from Theorem \ref{theorem:MIMO} by choosing $\Abf_1=\0bf$, $\Sigmabf_1=\Ibf-\Abf_2\Abf_2^T$ and $\Sigmabf_2=\Ibf$. Condition (\ref{eq:W1O1}) is automatically satisfied by $\Wbf_1\succeq\0bf=\Obf_1$.
\epf
\begin{remark}
The aligned-weak interference condition in \cite[Proposition 5]{Shang-etal:10IT_mimo} for the average power constraint is a special case of Proposition \ref{prop:MIMOZ}. The alignment weak interference means that if there exists a matrix $\Abf_2$ with $\Abf_2\Abf_2^T\preceq\Ibf$ and $\Fbf_2=\Abf_2^T\Hbf_2$, then treating interference as noise achieves the sum-rate capacity. Obviously, in such a case, all the conditions in Proposition \ref{prop:MIMOZ} are satisfied.

\end{remark}

In Sections \ref{section:MISO} and \ref{section:SIMO}, we apply Theorem \ref{theorem:MIMO} to MISO and SIMO channels and simplify the noisy-interference conditions.

\section{MISO ICs}
\label{section:MISO}
In \cite{Annapureddy&Veeravalli:09IT_submission}, it has been shown that the capacity of a two-user MISO IC is the same as that of a MISO IC with each transmitter having only two antennas. The main idea is to write the direct link channel vector as the sum of the interference channel vector and its orthogonal vector. The antenna reduction is also studied in \cite{Shang-etal:09IT_submission2} which shows that the single-user detection rate region of an $m-$user MISO IC with transmitter $i, 1\leq i\leq m$, having $t_i$ antennas, is the same as that of a MISO IC with transmitter $i$, having only $\min\{t_i,m\}$ antennas. The antenna reduction is performed systematically using \cite[eqs.(45)-(47)]{Shang-etal:09IT_submission2} which can also be used to show the equivalence of the capacity regions between the original $m$-user MISO IC and the new $m$-user MISO IC after antenna reduction. In the following, we apply the method in \cite{Shang-etal:09IT_submission2} to the two-user MISO IC to show the reduction process. On letting $\Hbf_i=\hat\hp_i^T$ and $\Fbf_i=\hat\fp_i^T$, $i=1,2$, in (\ref{eq:model}), the received signals of a MISO IC are
\bqa
Y_1&{}={}&\hat\hp_1^T\hat\xp_1+\hat\fp_2^T\hat\xp_2+Z_1\nn\\
Y_2&{}={}&\hat\hp_2^T\hat\xp_2+\hat\fp_1^T\hat\xp_1+Z_2
\label{eq:MISOmodelOld}
\eqa
where $\hbf_i$ and $\fbf_i$ are $t_i\times 1$ column vectors and we write the transmitted signal as $\hat\xp_i$ with power constraint $\hat P_i$. Define the singular value decomposition of $\fp_i$ as
\bqa
\hat\fp_i=\Ubf_i\left[\left\|\hat\fp\right\|,\0bf\right]^T
\label{eq:svd}
\eqa
where $\Ubf_i\Ubf_i^T=\Ibf$ and the dimension of the zero vector is $t_i-1$. Then we have
\bqa
\Ubf_i^T\hat\hp_i&{}\stackrel{(a)}={}&\left[\begin{array}{c}
                      \left\|\hat\hp_i\right\|\cos\theta_i \\
                      \gp_i
                    \end{array}
\right]\nn\\
&{}\stackrel{(b)}={}&\left[\begin{array}{cc}
                             1 &\quad \0bf \\
                             \0bf &\quad \Vbf_i
                           \end{array}
\right]\left[\begin{array}{c}
                      \left\|\hat\hp_i\right\|\cos\theta_i \\
                      \left\|\hat\hp_i\right\|\sin\theta_i\\
                      \0bf\\
                    \end{array}
\right]
\label{eq:Uh}
\eqa
where we define $\theta_i\triangleq\angle\left(\hat\hp_i,\hat\fp_i\right)$, and $\gp_i$ is a $(t_i-1)\times 1$ vector. Equality (a) follows from the fact that the first row of $\Ubf_i^T$ is $\hat\fp_i^T/\|\hat\fp_i\|$. Equality (b) is by the fact $\|\gp_i\|=\left\|\hat\hp_i\right\|\sin\theta_i$ and the singular value decomposition
\bqn
\gp_i=\Vbf_i\left[\left\|\hat\hp_i\right\|\sin\theta_i,\0bf\right]^T
\label{eq:gi}
\eqn
where $\Vbf_i^T\Vbf_i=\Ibf$, and the dimension of the zero vector is $t_i-2$.

Define
\bqa
\bar\xp\triangleq\Qbf_i\hat\xp_i
\eqa
where
\bqa
\Qbf_i=\left[\begin{array}{cc}
                             1 &\quad \0bf \\
                             \0bf &\quad \Vbf_i
                           \end{array}
\right]^T\Ubf_i^T.
\eqa
It is obvious that $\Qbf^T\Qbf=\Ibf$. Then the received signals of the MISO IC can be written as
\bqn
Y_1&{}={}&\left[\begin{array}{c}
                      \left\|\hat\hp_1\right\|\cos\theta_1 \\
                      \left\|\hat\hp_1\right\|\sin\theta_1\\
                      \0bf\\
                    \end{array}
\right]^T\bar\xp_1+\left[\begin{array}{c}
                      \left\|\hat\fp_2\right\| \\
                      0\\
                      \0bf\\
                    \end{array}
\right]^T\bar\xp_2+\zp_1\\
Y_2&{}={}&\left[\begin{array}{c}
                      \left\|\hat\hp_2\right\|\cos\theta_2 \\
                      \left\|\hat\hp_2\right\|\sin\theta_2\\
                      \0bf\\
                    \end{array}
\right]^T\bar\xp_2+\left[\begin{array}{c}
                      \left\|\hat\fp_1\right\| \\
                      0\\
                      \0bf\\
                    \end{array}
\right]^T\bar\xp_1+\zp_2.
\eqn
By removing irrelevant dimensions, we write the MISO IC in the following standard form:
\bqa
Y_1&{}={}&\hp_1^T\xp_1+\fp_2^T\xp_2+Z_1\nn\\
Y_2&{}={}&\hp_2^T\xp_2+\fp_1^T\xp_1+Z_2
\label{eq:MISOmodelNew}
\eqa
where the dimension of all the vectors is 2, and the power constraint for user $i$ is now $P_i$, and
\bqa
P_i&{}={}&\hat P_i\|\hat\hp_i\|^2\\
a_i&{}={}&\frac{\|\hat\fp_i\|^2}{\|\hat\hp_i\|^2}\\
\fp_i&{}={}&\left[\begin{array}{c}
                      \sqrt{a_i} \\
                      0\\
                    \end{array}
\right]\\
\hp_i&{}={}&\left[\begin{array}{c}
                      \cos\theta_i \\
                      \sin\theta_i\\
                     \end{array}
\right].
\eqa
Consequently, if $\Sbf_i$ is the input covariance matrix of user $i$ for equivalent channel (\ref{eq:MISOmodelNew}), the corresponding input covariance for the original channel is
\bqa
\hat\Sbf_i=\frac{1}{\left\|\hat\hp_i\right\|^2}\Qbf_i^T\left[\begin{array}{cc}
                             \Sbf_i &\quad \0bf \\
                             \0bf &\quad \0bf
                           \end{array}
\right]\Qbf_i.
\eqa

With the antenna reduction, we have the following result.
\begin{theorem}
For the MISO IC defined in (\ref{eq:MISOmodelOld}) and its equivalent channel (\ref{eq:MISOmodelNew}) with $\cos\angle\left(\hp_i,\fp_i\right)\neq 0$, $\fp_i\neq\0bf$, $\hp_i\neq\0bf$, $i=1,2$, denote $\Sbf_i^*$ as the optimal solution of problem (\ref{eq:lowerMIMO}) for the equivalent channel (\ref{eq:MISOmodelNew}), if $\Sbf_i^*\neq\0bf$ and
\bqa
&&\sigma_i^2\geq \bar\sigma_i^2,\quad i=1,2
\label{eq:MISOwioi}\\
&&\abs\left(A_1\right)+\abs\left(A_2\right)\leq 1
\label{eq:existence}
\eqa
where
\bqa
\sigma_1^2&{}={}&\frac{1}{2}\left[\left(1+A_1^2-A_2^2\right)+\sqrt{\left(1+A_1^2-A_2^2\right)^2-4A_1^2}\right]
\label{eq:sigma1MISO}\\
\sigma_2^2&{}={}&\frac{1}{2}\left[\left(1+A_2^2-A_1^2\right)+\sqrt{\left(1+A_2^2-A_1^2\right)^2-4A_2^2}\right]
\label{eq:sigma2MISO}\\
\bar\sigma_1^2&{}={}&-\fp_2^T\Sbf_2^*\fp_2+\frac{\sqrt{a_2}}{\cos\theta_2}\left(1+\hp_2^T\Sbf_2^*\hp_2+\fp_1^T\Sbf_1^*\fp_1\right)\frac{\fp_2^T\Sbf_2^*\hp_2}{\hp_2^T\Sbf_2^*\hp_2}
\label{eq:barSigma1MISO}\\
\bar\sigma_2^2&{}={}&-\fp_1^T\Sbf_1^*\fp_1+\frac{\sqrt{a_1}}{\cos\theta_1}\left(1+\hp_1^T\Sbf_1^*\hp_1+\fp_2^T\Sbf_2^*\fp_2\right)\frac{\fp_1^T\Sbf_1^*\hp_1}{\hp_1^T\Sbf_1^*\hp_1}
\label{eq:barSigma2MISO}\\
A_1&{}={}&\frac{\fp_1^T\Sbf_1^*\hp_1}{\hp_1^T\Sbf_1^*\hp_1}\left(1+\fp_2^T\Sbf_2^*\fp_2\right)
\label{eq:A1}\\
A_2&{}={}&\frac{\fp_2^T\Sbf_2^*\hp_2}{\hp_2^T\Sbf_2^*\hp_2}\left(1+\fp_1^T\Sbf_1^*\fp_1\right)
\label{eq:A2}
\eqa
then the sum-rate capacity is the maximum of problem (\ref{eq:lowerMIMO}) and is achieved by treating interference as noise.
\label{theorem:MISO}
\end{theorem}
\bpf
We use Theorem \ref{theorem:MIMO} to prove the converse. We first consider the existence of $\Abf_i$ (i.e., $\Abf_i=A_i$ in the MISO case) in (\ref{eq:Markov1}) and (\ref{eq:Markov2}) which require
\bqa
\Sbf_1^*\fp_1&{}={}&\Sbf_1^*\hp_1\left(1+\fp_2^T\Sbf_2^*\fp_2\right)^{-1}A_1
\label{eq:MCmiso1}\\
\Sbf_2^*\fp_2&{}={}&\Sbf_2^*\hp_2\left(1+\fp_1^T\Sbf_1^*\fp_1\right)^{-1}A_2.
\label{eq:MCmiso2}
\eqa
It has been shown in \cite{Shang-etal:09IT_submission2} that $\rank\left(\Sbf_i^*\right)\leq 1$. With the assumption $\tr(\Sbf_i^*)>0$, we have
\bqa
\rank\left(\Sbf_i^*\right)=1.
\label{eq:unitRank}
\eqa
Then we can write
\bqa
\Sbf_i^*=\gammabf_i\gammabf_i^T
\eqa
where $\gammabf$ is a $2\times 1$ vector. We have
\bqa
\gammabf_1\gammabf_1^T\fp_1&{}={}&\gammabf_1\gammabf_1^T\hp_1\left(1+\fp_2^T\Sbf_2^*\fp_2\right)^{-1}A_1.
\eqa
Obviously, if $\gammabf^T\hp_1=0$, then $\gammabf^T\fp_1=0$ because otherwise transmitter $1$ does not transmit anything to receiver $1$ while still generating interference to receiver $2$. In this case $A_1$ can choose any value. If $\gammabf^T\hp_1\neq 0$, we have
\bqa
A_1=\frac{\gammabf_1^T\fp_1}{\gammabf_1^T\hp_1\left(1+\fp_2^T\Sbf_2^*\fp_2\right)^{-1}}.
\eqa
Therefore, $\Abf_1$ always exists. Similarly, we can show the existence of $A_2$. Another expression of $A_i$ in (\ref{eq:A1}) and (\ref{eq:A2}) is obtained by left-multiply (\ref{eq:MCmiso1}) and (\ref{eq:MCmiso2}) with $\hp_1^T$ and $\hp_2^T$, respectively.

We then consider the existence of $\Sigmabf_i$ (i.e., $\Sigmabf_i=\sigma_i^2$ in the MISO case) in (\ref{eq:sigma1}) and (\ref{eq:sigma2}). By choosing equality in both (\ref{eq:sigma1}) and (\ref{eq:sigma2}), we obtain $\sigma_i^2$ in (\ref{eq:sigma1MISO}) and (\ref{eq:sigma2MISO}). It can be shown that the existence of $\sigma_i^2$, or equivalently, that (\ref{eq:sigma1MISO}) and (\ref{eq:sigma2MISO}) are feasible, is guaranteed by (\ref{eq:existence}) (details can be found in \cite[p. 696]{Shang-etal:09IT}).

It remains to consider whether conditions (\ref{eq:W1O1}) and (\ref{eq:W2O2}) are satisfied. In the following, we do not verify these two conditions directly from  (\ref{eq:W1}) or (\ref{eq:W2}). Instead, we use the equivalent conditions (\ref{eq:lowerKKTS1})-(\ref{eq:lowerIneq}) since we have additional information (\ref{eq:unitRank}) for $\Sbf_i^*$.

From (\ref{eq:lowerIneq}), the columns of $\Wbf_i$ are all in the eigenvector space of $\Sbf_i^*$ associated with its zero eigenvalue. Since $\rank\left(\Sbf_i^*\right)=1$ and $\Sbf_i^*$ is a $2\times 2$ matrix, the dimension of this eigenvector space is $1$. By (\ref{eq:MCmiso1}), the eigenvector is $A_1\left(1+\fp_2^T\Sbf_2^*\fp_2\right)^{-1}\hp_1-\fp_1$. Therefore, there exist a constant $k\geq 0$ such that
\bqa
\Wbf_1=k\left(A_1\left(1+\fp_2^T\Sbf_2^*\fp_2\right)^{-1}\hp_1-\fp_1\right)\left(A_1\left(1+\fp_2^T\Sbf_2^*\fp_2\right)^{-1}\hp_1-\fp_1\right)^T.
\label{eq:W1misoExp1}
\eqa
On the other hand, from (\ref{eq:lowerKKTS1}) we have
\bqa
\Wbf_1=-\frac{\hp_1\hp_1^T}{2\left(1+\hp_1^T\Sbf_1^*\hp_1+\fp_2^T\Sbf_2^*\fp_2\right)}+\frac{\hp_2^T\Sbf_2^*\hp_2\cdot\fp_1\fp_1^T}
{2\left(1+\fp_1^T\Sbf_1^*\fp_1\right)\left(1+\hp_2^T\Sbf_2^*\hp_2+\fp_1^T\Sbf_1^*\fp_1\right)}+\lambda_1\Ibf.
\label{eq:W1misoExp2}
\eqa
On comparing the element of $\Wbf_1$ on the first row and the second column in expression (\ref{eq:W1misoExp1}) and (\ref{eq:W1misoExp2}), we have
\bqa
k=\frac{-\cos\theta_1}{2\left(1+\hp_1^T\Sbf_1^*\hp_1+\fp_2^T\Sbf_2^*\fp_2\right)\dfrac{\fp_1^T\Sbf_1^*\hp_1}{\hp_1^T\Sbf_1^*\hp_1}
\left(\dfrac{\fp_1^T\Sbf_1^*\hp_1}{\hp_1^T\Sbf_1^*\hp_1}\cos\theta_1-\sqrt{a_1}\right)}.
\eqa
From (\ref{eq:Obf1}), we have
\bqa
O_1=\frac{1}{2}\frac{\left(A_1\left(1+\fp_2^T\Sbf_2^*\fp_2\right)^{-1}\hp_1-\fp_1\right)\left(A_1\left(1+\fp_2^T\Sbf_2^*\fp_2\right)^{-1}\hp_1-\fp_1\right)^T}
{\sigma_1^2-\dfrac{A_1^2}{1+\fp_2\Sbf_2^*\fp_2}}.
\eqa
Therefore, condition (\ref{eq:W1O1}) requires
\bqa
k\geq\frac{1}{\sigma_1^2-\dfrac{A_1^2}{1+\fp_2\Sbf_2^*\fp_2}}
\eqa
which is equivalent to (\ref{eq:MISOwioi}). Similarly, (\ref{eq:MISOwioi}) guarantees that (\ref{eq:W2O2}) is satisfied. Therefore, under conditions (\ref{eq:MISOwioi}) and (\ref{eq:existence}), all the requirements of Theorem \ref{theorem:MIMO} are satisfied, and the MISO IC has noisy interference.
\epf
\begin{remark}
Consider the computation of the noisy-interference sum-rate capacity of a MISO IC. Using \cite[Theorem 1]{Shang-etal:09IT_submission2}, the maximum of problem (\ref{eq:lowerMIMO}) is
\bqa
\max_{\phi_i\in\left[0,\textrm{abs}\left(\frac{\pi}{2}-\theta_i\right)\right]}
\frac{1}{2}\log\left(1+\frac{P_1\sin^2\left(\theta_1+\rho_1\phi_1\right)}{1+a_2P_2\sin^2\phi_2}\right)
+\frac{1}{2}\log\left(1+\frac{P_2\sin^2\left(\theta_2+\rho_2\phi_2\right)}{1+a_1P_1\sin^2\phi_1}\right)
\label{eq:MISObfs}
\eqa
where $\rho_i=1$ if $\theta_i\in\left[0,\frac{\pi}{2}\right]$ and $\rho_i=-1$ otherwise. If $\phi_i^*$ is optimal, then the corresponding input covariance matrix is
\bqa
\Sbf_i^*=P_i\left[\begin{array}{cc}
                    \sin^2\phi_i^* &\quad \rho_i\sin\phi_i^*\cos\phi_i^* \\
                    \rho_i\sin\phi_i^*\cos\phi_i^* &\quad \cos^2\phi_i^*
                  \end{array}
\right].
\label{eq:MISOSstar}
\eqa
A closed-form expression for $\phi_i^*$ is difficult to obtain for the general MISO ICs, or even MISO ZICs. However, if the MISO IC is symmetric with $\theta_1=\theta_2=\theta$, $a_1=a_2=a$ and $P_1=P_2=P$, then we have:
\bqa
\tan\phi^*=\textrm{abs}\left(\frac{1}{(1+aP)\tan\theta}\right).
\label{eq:optPhi}
\eqa
\end{remark}

\begin{remark}
If the MISO IC is symmetric as defined above, the noisy-interference condition is given in \cite[Theorem 2]{Annapureddy&Veeravalli:09IT_submission}, which can also be obtained from Theorem \ref{theorem:MISO}. In this case, the optimal $\Sbf_i^*$ is given in  (\ref{eq:MISOSstar}) and (\ref{eq:optPhi}). Conditions in Theorem \ref{theorem:MISO} reduce to
\bqa
&&A=\frac{\fp\Sbf^*\hp}{\hp^T\Sbf^*\hp}\left(1+\fp^T\Sbf^*\fp\right)\leq\frac{1}{2}\\
&&\sigma^2=\frac{1}{2}+\frac{1}{2}\sqrt{1-4A^2}\leq\bar\sigma^2=-\fp^T\Sbf^*\fp+\frac{\sqrt{a}}{\cos\theta}\left(\fp^T\Sbf^*\fp+A\right).
\eqa
The above conditions are exactly \cite[eq.(53)]{Annapureddy&Veeravalli:09IT_submission} which are satisfied under the conditions in \cite[Theorem 2]{Annapureddy&Veeravalli:09IT_submission}.

\end{remark}

Theorem \ref{theorem:MISO} applies to the case in which $\cos\theta_i\neq 0$ and $\|\fp_i\|\neq 0$. If any of these two conditions are satisfied, the MISO IC reduces to a MISO ZIC. The noisy-interference sum-rate capacity is obtain in the following proposition.

\begin{proposition}
For the MISO IC defined in (\ref{eq:MISOmodelOld}) and its equivalent channel (\ref{eq:MISOmodelNew}) with $\cos\angle\left(\hp_1,\fp_1\right)=\frac{\pi}{2}$, or $\fp_1=\0bf$, denote by $\Sbf_i^*$, $i=1,2$, the optimal solution of problem (\ref{eq:lowerMIMO}) for the equivalent channel (\ref{eq:MISOmodelNew}). If $\Sbf_i^*\neq\0bf$ and
\bqa
&&\fp_2^T\Sbf_2^*\fp_2\leq\hp_2^T\Sbf_2^*\hp_2
\label{eq:MISOZcond1}\\
&&a_2\left[\frac{\fp_2^T\Sbf_2^*\hp_2\left(1+\hp_2^T\Sbf_2^*\hp_2\right)}{\hp_2^T\Sbf_2^*\hp_2\left(1+\fp_2^T\Sbf_2^*\fp_2\right)}\right]^2
\leq \cos^2\theta_2
\label{eq:MISOZcond2}
\eqa
then the sum-rate capacity is the maximum in problem (\ref{eq:lowerMIMO}) and is achieved by treating interference as noise.
\label{prop:MISOZ}
\end{proposition}
\bpf
We first consider the case when $\fp_1=\0bf$. From (\ref{eq:sigma1MISO})-(\ref{eq:A2}), we have
\bqa
\sigma_1^2&{}={}&1-A_2^2\nn\\
\sigma_2^2&{}={}&1\nn\\
\bar\sigma_1^2&{}={}&-\fp_2\Sbf_2^*\fp_2+\frac{\sqrt{a_2}}{\cos\theta_2}\left(1+\hp_2^T\Sbf_2^*\hp_2\right)\frac{\fp_2^T\Sbf_2^*\hp_2}{\hp_2^T\Sbf_2^*\hp_2}\nn\\
\bar\sigma_2^2&{}={}&0\nn\\
A_1&{}={}&0\nn\\
A_2&{}={}&\frac{\fp_2^T\Sbf_2^*\hp_2}{\hp_2^T\Sbf_2^*\hp_2}.\nn
\eqa
Condition (\ref{eq:MISOZcond1}) guarantees that (\ref{eq:existence}) is satisfied since
\bqa
A_2^2=\left(\frac{\fp_2^T\Sbf_2^*\hp_2}{\hp_2^T\Sbf_2^*\hp_2}\right)^2=\frac{\fp_2^T\Sbf_2^*\fp_2}{\hp_2^T\Sbf_2^*\hp_2}
\eqa
due to the fact that $\rank \left(\Sbf_2^*\right)=1$. Then it remains to consider (\ref{eq:MISOwioi}) for $i=1$, which is satisfied by (\ref{eq:MISOZcond2}) on the condition
\bqa
\frac{\fp_2^T\Sbf_2^*\hp_2}{\cos\theta_2}\geq 0\nn
\eqa
which is true by (\ref{eq:MISOSstar}):
\bqn
\frac{\fp_2^T\Sbf_2^*\hp_2}{\cos\theta_2}=\sqrt{a_2}P_2\frac{\sin^2\phi_2^*\cos\theta_2+\rho_2\sin\phi_2^*\cos\phi_2^*\sin\theta_2}{\cos\theta_2}\geq 0.
\eqn

In the case $\fp_1\neq\0bf$ but $\theta_1=\frac{\pi}{2}$, the capacity region is outer bound by that of the same channel but with $\fp_1=\0bf$. If (\ref{eq:MISOZcond1}) and (\ref{eq:MISOZcond2}) are satisfied, then the sum-rate capacity of the channel with $\fp_1=\0bf$ is an outer bound on that of the channel with $\fp_1\neq\0bf$ but $\theta_1=\frac{\pi}{2}$. The achievability is due to the fact that
\bqa
\fp_1^T\Sbf_1^*\fp_1=0\nn
\eqa
since
\bqa
\Sbf_1^*=P_1\hp_1\hp_1^T.\nn
\eqa
We note that Proposition \ref{prop:MISOZ} can also be proved by Proposition \ref{prop:MIMOZ}.
\epf

\section{SIMO ICs}
\label{section:SIMO}

On letting $\Hbf_i=\hat\hp_i$ and $\Fbf_i=\hat\fp_i$, $i=1,2$, in (\ref{eq:model}), the received signals of a MISO IC are
\bqa
\hat\yp_1&{}={}&\hat\hp_1 X_1+\hat\fp_2 X_2+\hat\zp_1\nn\\
\hat\yp_2&{}={}&\hat\hp_2 X_2+\hat\fp_1 X_1+\hat\zp_2
\label{eq:SIMOmodelOld}
\eqa
where $\hbf_i$ and $\fbf_i$ are $t_i\times 1$ vectors and we write the transmitted signal as $\hat\xp_i$ with power constraint $\hat P_i$.

We can follow the same process (\ref{eq:svd})-(\ref{eq:gi}) in Section \ref{section:MISO} to find the equivalent channel for (\ref{eq:SIMOmodelOld}) with reduced number of antennas. The difference is that we need to replace the $\hp_i$ in (\ref{eq:Uh}) with $\hp_j$ where $j\neq i$. Then we left-multiply $\yp_i$ with $\Qbf_i$ and obtain the equivalent channel
\bqa
\yp_1&{}={}&\hp_1 X_1+\fp_2 X_2+\zp_1\nn\\
\yp_2&{}={}&\hp_2 X_2+\fp_1 X_1+\zp_2
\label{eq:SIMOmodelNew}
\eqa
where the dimension of all the vectors is 2, the power constraint for user $i$ is now $P_i$, and
\bqa
P_i&{}={}&\hat P_i\\
a_i&{}={}&\frac{\|\hat\fp_i\|^2}{\|\hat\hp_i\|^2}\\
\fp_i&{}={}&\left[\begin{array}{c}
                      \sqrt{a_i} \\
                      \0bf\\
                    \end{array}
\right]\\
\hp_i&{}={}&\left[\begin{array}{c}
                      \cos\varphi_i \\
                      \sin\varphi_i\\
                     \end{array}
\right]\\
\varphi_i&{}={}&\angle\left(\hp_i,\fp_j\right)\qquad i,j\in\{1,2\}, j\neq i.
\eqa

We first present the noisy-interference sum-rate capacity of the SIMO ZIC as this is a special case of \cite[Proposition 5]{Shang-etal:10IT_mimo}.
\begin{proposition}
\cite[Proposition 5]{Shang-etal:10IT_mimo} For the SIMO IC defined in (\ref{eq:SIMOmodelOld}) and its equivalent channel (\ref{eq:SIMOmodelNew}) with $\varphi_2=\frac{\pi}{2}$ or $\fp_1=0$, if $\|\fp_2\|\leq\|\hp_2\|$, then the sum-rate capacity is
\bqa
\frac{1}{2}\log\left|\Ibf+P_1\hp_1\hp_1^T\left(\Ibf+P_2\fp_2\fp_2^T\right)^{-1}\right|+\frac{1}{2}\log\left|\Ibf+P_2\hp_2\hp_2^T\right|.
\eqa
\label{prop:SIMOZ}
\end{proposition}
\bpf
We first consider the case when $\fp_1=0$. Then from \cite[Proposition 5]{Shang-etal:10IT_mimo}, if there exists a matrix $\Abf_2$ such that
\bqa
\fp_2&{}={}&\Abf_2^T\hp_2
\label{eq:SIMOZweak}\\
\Ibf&{}\succeq{}&\Abf_2^T\Abf_2
\label{eq:SIMOZA}
\eqa
then the sum-rate capacity is
\bqa
\max_{0\leq S_i\leq P_i,i=1,2}\frac{1}{2}\log\left|\Ibf+S_1\hp_1\hp_1^T\left(\Ibf+S_2\fp_2\fp_2^T\right)^{-1}\right|+\frac{1}{2}\log\left|\Ibf+S_2\hp_2\hp_2^T\right|.
\label{eq:optSIMOZ}
\eqa
Then we can choose
\bqa
\Abf_2^T=\frac{\fp_2\hp_2^T}{\|\hp_2\|^2}=\fp_2\hp_2^T
\eqa
and (\ref{eq:SIMOZweak}) is satisfied. For (\ref{eq:SIMOZA}), we observe
\bqa
\Abf_2^T\Abf_2=\fp_2\left(\hp_2^T\hp_2\right)\fp_2^T=\left[\begin{array}{cc}
                                                             a_2 &\quad 0 \\
                                                             0 &\quad 0
                                                           \end{array}
\right]\preceq\Ibf
\label{eq:AAleqI}
\eqa
where the last equality is by the assumption $\|\fp_2\|\leq\|\hp_2\|$.

Then we need to show that $S_i^*=P_i$ maximizes (\ref{eq:optSIMOZ}). On denoting the objective function of (\ref{eq:optSIMOZ}) by $R_s$, we have
\bqa
\frac{\partial R_s}{\partial S_1}=\frac{1}{2}\hp_1^T\left(\Ibf+S_1\hp_1\hp_1^T+S_2\fp_2\fp_2^T\right)^{-1}\hp_1\geq 0
\eqa
and
\bqa
&&\frac{\partial R_s}{\partial S_2}\nn\\
&&=\frac{1}{2}\fp_2^T\left(\Ibf+S_1\hp_1\hp_1^T+S_2\fp_2\fp_2^T\right)^{-1}\fp_2
-\frac{1}{2}\fp_2^T\left(\Ibf+S_2\fp_2\fp_2^T\right)^{-1}\fp_2+\frac{1}{2}\hp_2^T\left(\Ibf+S_2\hp_2\hp_2^T\right)^{-1}\hp_2\nn\\
&&\geq-\frac{1}{2}\fp_2^T\left(\Ibf+S_2\fp_2\fp_2^T\right)^{-1}\fp_2+\frac{1}{2}\hp_2^T\left(\Ibf+S_2\hp_2\hp_2^T\right)^{-1}\hp_2\nn\\
&&\stackrel{(a)}=-\frac{1}{2}\left(1+S_2\fp_2^T\fp_2\right)^{-1}\fp_2^T\fp_2+\frac{1}{2}\left(1+S_2\hp_2^T\hp_2\right)^{-1}\hp_2^T\hp_2\nn\\
&&=\frac{\|\hp_2\|^2-\|\fp_2\|^2}{2\left(1+S_2\|\fp_2\|^2\right)\left(\Ibf+S_2\|\hp_2\|^2\right)}\nn\\
&&\geq 0
\eqa
where (a) is by the matrix identity (\ref{eq:Searle1}). Therefore $R_s$ is maximized by $S_i^*=P_i$.

In the case when $\fp_1\neq\0bf$ and $\varphi_2=\frac{\pi}{2}$, the converse can be proved by assuming $\fp_1=0$ to eliminate the interference, and the achievability is proved by left-multiplying $\yp_2$ with $\hp_2$ to null out the interference.

We note that Proposition \ref{prop:SIMOZ} can also be proved by Proposition \ref{prop:MIMOZ}.
\epf
\begin{theorem}
For the SIMO IC defined in (\ref{eq:SIMOmodelOld}) and its equivalent channel (\ref{eq:SIMOmodelNew}), if for $i=1$ or $2$
\bqa
\textrm{radius}\left(\Phibf_i\right)\leq\frac{1}{2}
\label{eq:SIMOcond}
\eqa
where
\bqa%
&&\Phibf_1=\left(\Ibf-\Abf_1^T\Abf_1-\Abf_2\Abf_2^T\right)^{-\frac{1}{2}}\Abf_1^T\Abf_2^T
\left(\Ibf-\Abf_1^T\Abf_1-\Abf_2\Abf_2^T\right)^{-\frac{1}{2}}
\label{eq:SIMOphi1}\\
&&\Phibf_2=\left(\Ibf-\Abf_1\Abf_1^T-\Abf_2^T\Abf_2\right)^{-\frac{1}{2}}\Abf_2^T\Abf_1^T
\left(\Ibf-\Abf_1\Abf_1^T-\Abf_2^T\Abf_2\right)^{-\frac{1}{2}}
\label{eq:SIMOphi2}\\
&&\Abf_1\left(\Ibf+P_2\fp_2\fp_2^T\right)\hp_1=\fp_1
\label{eq:SIMOA1}\\
&&\Abf_2\left(\Ibf+P_1\fp_1\fp_1^T\right)\hp_2=\fp_2
\label{eq:SIMOA2}
\eqa
then the sum-rate capacity is
\bqa
\frac{1}{2}\log\left|\Ibf+P_1\hp_1\hp_1^T\left(\Ibf+P_2\fp_2\fp_2^T\right)^{-1}\right|
+\frac{1}{2}\log\left|\Ibf+P_2\hp_2\hp_2^T\left(\Ibf+P_1\fp_1\fp_1^T\right)^{-1}\right|.
\eqa
\label{theorem:SIMO}
\end{theorem}

\bpf
We prove Theorem \ref{theorem:SIMO} from Theorem \ref{theorem:upper} instead of Theorem \ref{theorem:MIMO} since the optimal solution is known for problem (\ref{eq:upper}). If we choose $\Abf_i$ in (\ref{eq:SIMOA1}) and (\ref{eq:SIMOA2}), then by Lemma \ref{lemma:myMxInv}, given (\ref{eq:SIMOcond}) there exists $\Sigmabf_i$ such that
\bqa
&&\Abf_1^T\Abf_1\prec\Sigmabf_1=\Ibf-\Abf_2\Sigmabf_2^{-1}\Abf_2^T
\label{eq:ricatti1}\\
&&\Abf_2^T\Abf_2\prec\Sigmabf_2=\Ibf-\Abf_1\Sigmabf_1^{-1}\Abf_1^T.
\label{eq:ricatti2}
\eqa
Therefore, conditions (\ref{eq:Ei})-(\ref{eq:sigma2}) are satisfied. In the following, we show that the upper bound $R_{1u}(S_1,S_2)+R_{2u}(S_1,S_2)$ is maximized at $S_i^*=P_i$ and $R_{1u}(P_1,P_2)+R_{2u}(P_1,P_2)=R_{1l}(P_1,P_2)+R_{2l}(P_1,P_2)$.

From (\ref{eq:upperMutualI}) we have
\bqa
&&R_{1u}+R_{2u}\nn\\
&&=I\left( X_{1G};\left[\begin{array}{c}
                                                    \hp_1 \\
                                                    \fp_1
                                                  \end{array}
\right] X_{1G}+\left[\begin{array}{c}
                                                    \fp_2 \\
                                                    \0bf
                                                  \end{array}
\right] X_{2G}+\left[\begin{array}{c}
                                                    \zp_1 \\
                                                    \np_1
                                                  \end{array}
\right]\right)+I\left( X_{1G};\left[\begin{array}{c}
                                                    \hp_2 \\
                                                    \fp_2
                                                  \end{array}
\right] X_{2G}+\left[\begin{array}{c}
                                                    \fp_1 \\
                                                    \0bf
                                                  \end{array}
\right] X_{1G}+\left[\begin{array}{c}
                                                    \zp_2 \\
                                                    \np_2
                                                  \end{array}
\right]\right)\nn\\
&&=h\left(\fp_1 X_{1G}+\np_1\right)-h\left(\np_1\right)+h\left(\hp_1 X_{1G}+\fp_2 X_{2G}+\zp_1|\fp_1 X_{1G}+\np_1\right)-h\left(\fp_2 X_{2G}+\zp_1|\np_1\right)\nn\\
  &&\hspace{-.1in}\hspace{.2in}+h\left(\fp_2 X_{2G}+\np_2\right)-h\left(\np_2\right)+h\left(\hp_2 X_{2G}+\fp_1 X_{1G}+\zp_2|\fp_2 X_{2G}+\np_2\right)-h\left(\fp_1 X_{1G}+\zp_2|\np_2\right)
\nn\\
&&=-h\left(\np_1\right)+h\left(\hp_1 X_{1G}+\fp_2 X_{2G}+\zp_1|\fp_1 X_{1G}+\np_1\right)-h\left(\np_2\right)+h\left(\hp_2 X_{2G}+\fp_1 X_{1G}+\zp_2|\fp_2 X_{2G}+\np_2\right)\nn\\
\eqa
where the last equality is by (\ref{eq:ricatti1}) and (\ref{eq:ricatti2}) which mean
\bqa
\Cov(\np_i)=\Cov(\zp_j|\np_j)\qquad i,j\in\{1,2\}, i\neq j.
\eqa
Then it suffices to show that $h\left(\hp_1 X_{1G}+\fp_2 X_{2G}+\zp_1|\fp_1 X_{1G}+\np_1\right)$ is an increasing function of $\Cov(X_{iG})$. We write $X_{iG}=\bar X_{iG}+\hat X_{iG}$ where $X_{iG}$ and $\hat X_{iG}$ are independent Gaussian variables. Obviously, we have $\Cov(X_{iG})\geq \Cov(\bar X_{iG})$ and
\bqa
&&h\left(\hp_1 X_{1G}+\fp_2 X_{2G}+\zp_1|\fp_1 X_{1G}+\np_1\right)\nn\\
&&\geq h\left(\hp_1 X_{1G}+\fp_2 X_{2G}+\zp_1|\fp_1 X_{1G}+\np_1,\hat X_{1G},\hat X_{2G}\right)\nn\\
&&=h\left(\hp_1\bar X_{1G}+\fp_2\bar X_{2G}+\zp_1|\fp_1\bar X_{1G}+\np_1\right).
\eqa
Therefore,  the upper bound $R_{1u}(S_1,S_2)+R_{2u}(S_1,S_2)$ is maximized at $S_i^*=P_i$. From (\ref{eq:R1u}), (\ref{eq:R2u}), (\ref{eq:SIMOA1}) and (\ref{eq:SIMOA2}), we have
\bqa
R_{iu}(P_1,P_2)=R_{il}(P_1,P_2).
\eqa
Therefore, the upper bound is achievable and hence is the sum-rate capacity.
\epf
\begin{remark}
A simple way to choose matrix $\Abf_i$ that satisfies (\ref{eq:SIMOA1}) and (\ref{eq:SIMOA2}) is to let
\bqa
\Abf_1&{}={}&\left(\Ibf+P_2\fp_2\fp_2^T\right)\hp_1\fp_1^T
\label{eq:simoA1simple}\\
\Abf_2&{}={}&\left(\Ibf+P_1\fp_1\fp_1^T\right)\hp_2\fp_2^T.
\label{eq:simoA2simple}
\eqa
However, this may not always be the best choice for (\ref{eq:SIMOcond}). An alternative way is to let \cite[eq. (39)]{Annapureddy&Veeravalli:09IT_submission}
\bqa
\Abf_i=\frac{\vp_i\fp_i^T}{\hp_i^T\left(1+P_j\fp_j\fp_j^T\right)\vp_i}
\eqa
where  $\vp_i$ is a vector. Then, to satisfy (\ref{eq:SIMOcond}), we need only
\bqa
\min_{\vp_1,\vp_2}\textrm{radius}(\Phibf_i)\leq\frac{1}{2}.
\eqa
\end{remark}
\begin{remark}
Proposition \ref{prop:SIMOZ} can also be obtained from Theorem \ref{theorem:SIMO}. Let $\fp_1=\0bf$, then we have $\Abf_1=\0bf$, $\Abf_2=\hp_2\fp_2^T$ and $\Phibf_i=\0bf$. Therefore, condition (\ref{eq:SIMOcond}) is always satisfied. Notice that $\left(\Ibf-\Abf_2\Abf_2^T\right)^{-\frac{1}{2}}$ and $\left(\Ibf-\Abf_2^T\Abf_2\right)^{-\frac{1}{2}}$ must exist such that $\Phibf_i$ exists. By \cite[Lemma 7]{Shang-etal:10IT_mimo} this requires $\Abf_2^T\Abf_2\preceq\Ibf$ which is (\ref{eq:AAleqI}).
\end{remark}
\begin{remark}
If the SIMO IC is symmetric, i.e., $\hp_1=\hp_2=\hp$, $\fp_1=\fp_2=\fp$ and $P_1=P_2=P$, the noisy-interference condition is given in \cite[Theorem 3]{Annapureddy&Veeravalli:09IT_submission}. We will show that the same result can be obtained from Theorem \ref{theorem:SIMO}. Without loss of generality, we assume $\theta\in\left[0,\frac{\pi}{2}\right]$. The matrix $\Abf$ that satisfies (\ref{eq:SIMOA1}) and (\ref{eq:SIMOA2}) can be chosen as
\bqa
\Abf=\frac{\sqrt{a}}{\dfrac{\cos\omega\cos\theta}{1+aP}+\sin\omega\sin\theta}\left[\begin{array}{cc}
    \cos\omega&\quad\sin\omega\\
    0&\quad 0
                                                                                   \end{array}
\right]
\eqa
where $\omega$ is a real number. Since $\Abf_1=\Abf_2$, condition (\ref{eq:SIMOcond}) reduces to $\textrm{radius}(\Abf)\leq \frac{1}{2}$, i.e.,
\bqa
\frac{1}{2}&{}\geq{}&\min_{\omega}\hspace{.05in}\max_{\phi}\hspace{.05in}\abs\left(\left[\begin{array}{c}
                                                                                           \cos\phi \\
                                                                                           \sin\phi
                                                                                         \end{array}
\right]^T\Abf\left[\begin{array}{c}
                                                                                           \cos\phi \\
                                                                                           \sin\phi
                                                                                         \end{array}
\right]\right)\nn\\
&{}={}&\min_{\omega}\hspace{.05in}\max_{\phi}\hspace{.05in}\abs\left[\frac{\sqrt{a}\left(\cos^2\phi\cos\omega+\cos\phi\sin\phi\sin\omega\right)}
{\dfrac{\cos\omega\cos\theta}{1+aP}+\sin\omega\sin\theta}\right]\nn\\
&{}={}&\min_{\omega}\frac{\sqrt{a}\left(1+\abs(\cos\omega)\right)/2}{\abs\left[\dfrac{\cos\omega\cos\theta}{1+aP}+\sin\omega\sin\theta\right]}\nn\\
&{}={}&\min_{\omega\in\left[0,\frac{\pi}{2}\right]}\frac{\sqrt{a}\left(1+\cos\omega\right)/2}{\sqrt{r}\sin(\omega+\beta)}
\label{eq:SIMOradius}
\eqa
where
\bqa
r&{}={}&\frac{\cos^2\theta}{(1+aP)^2}+\sin^2\theta\nn\\
\beta&{}={}&\textrm{atan}\frac{\cos\theta}{(1+aP)\sin\theta}\in\left[0,\frac{\pi}{2}\right].
\eqa
It can be shown that the optimal $\omega$ for (\ref{eq:SIMOradius}) is
\bqa
\omega=\left\{\begin{array}{cl}
                \frac{\pi}{2}, & \quad\textrm{if }\beta\in\left[0,\frac{\pi}{4}\right] \\
                \pi-2\beta, & \quad\textrm{if } \beta\in\left[\frac{\pi}{4},\frac{\pi}{2}\right].
              \end{array}
\right.
\eqa
Then (\ref{eq:SIMOradius}) becomes
\bqa
  &&a\leq\sin^2\theta \hspace{2in} \textrm{if } \frac{\cos\theta}{(1+aP)}\leq\sin\theta\\
  &&\frac{\cos^2\theta}{(1+aP)^2}-\frac{2\sqrt{a}\cos\theta}{1+aP}+\sin^2\theta\geq 0 \qquad \textrm{otherwise}
\eqa
which are exactly the conditions in \cite[Theorem 3]{Annapureddy&Veeravalli:09IT_submission}.
\label{remark:sSIMO}
\end{remark}

\section{Numerical examples}
\label{section:example}

\begin{example}
Consider a MIMO IC with channel matrices:
\bqn
&&\Hbf_1=\left[\begin{array}{cc}
                    -1.4510&\quad   -1.0078\\
                    -1.8953&\quad    0.2184\\
                    1.9125&\quad   -1.6068
             \end{array}
\right],\quad
\Fbf_2=\left[\begin{array}{ccc}
                0.4255&\quad   -0.1702&\quad    0.6865\\
                0.5133&\quad    0.1574&\quad    0.1805\\
                -0.4795&\quad   -0.5019&\quad    0.4648
             \end{array}
\right],\\
&&\Hbf_2=\left[\begin{array}{ccc}
                0.7739&\quad    1.4112&\quad   -1.8231\\
                1.4817 &\quad  -0.4647&\quad    2.1620
             \end{array}
\right]\quad\textrm{and}\quad
\Fbf_1=\left[\begin{array}{cc}
                -0.2636&\quad    0.2981\\
                -0.3483&\quad   -0.1426
             \end{array}
\right]
\eqn
and power constraints:
\bqn
P_1=1 \quad\textrm{and } P_2=4.
\eqn

The optimal input covariance matrices for problem (\ref{eq:lowerMIMO}) are
\bqn
\Sbf_1^*=\left[\begin{array}{cc}
                0.9079&\quad    -0.2892\\
                -0.2892&\quad     0.0921
               \end{array}
\right]\quad\textrm{and}\quad
\Sbf_2^*=\left[\begin{array}{ccc}
                0.9458&\quad    0.1788&\quad    0.5314\\
                0.1788&\quad    0.6839&\quad   -1.0601\\
                0.5314&\quad   -1.0601&\quad    2.3703
               \end{array}
\right]
\eqn
and both $\Sbf_1^*$ and $\Sbf_2^*$ are singular:
\bqn
\rank(\Sbf_1^*)=1\quad\textrm{and}\quad\rank(\Sbf_2^*)=2.
\eqn
The $\Gbf_1$ and $\Gbf_2$ in (\ref{eq:lowerKKTS1}) and (\ref{eq:lowerKKTS2}) and the Lagrangian multipliers are
\bqn
&&\Gbf_1=\left[\begin{array}{cc}
                -0.3624&\quad    0.0005\\
                0.0005&\quad   -0.3608
             \end{array}
\right],\quad
\Gbf_2=\left[\begin{array}{ccc}
                -0.1368&\quad   -0.0525&\quad   -0.0294\\
                -0.0525&\quad   -0.0591&\quad    0.0583\\
                -0.0294&\quad    0.0583&\quad   -0.1305
             \end{array}
\right]\\
&&\Wbf_1=\left[\begin{array}{cc}
                0.1740&\quad    0.5463\\
                0.5463&\quad    1.7150
             \end{array}
\right]*10^{-3},\quad
\Wbf_2=\left[\begin{array}{ccc}
                2.6419&\quad   -5.2450&\quad   -2.9381\\
                -5.2450&\quad   10.4117&\quad    5.8325\\
                 -2.9381&\quad    5.8325&\quad    3.2674
             \end{array}
\right]*10^{-2}\\
&&\lambda_1= 0.3626\quad\textrm{and}\quad\lambda_2=0.1632.
\eqn
It is easy to verify that the KKT conditions in (\ref{eq:lowerKKTS1})-(\ref{eq:lowerIneq}) are satisfied.

The $\Abf_1$ and $\Abf_2$ that satisfy (\ref{eq:Markov1}) and (\ref{eq:Markov2}) are
\bqn
\Abf_1=\left[\begin{array}{cc}
                -0.2821&\quad    0.4705\\
                0.0254&\quad    0.2073\\
                -0.3814&\quad    0.1588
             \end{array}
\right]\quad\textrm{and}\quad
\Abf_2=\left[\begin{array}{ccc}
                0.0047&\quad    0.2392&\quad   -0.4520\\
                0.3215&\quad    0.2853&\quad   -0.1663
             \end{array}
\right].
\eqn
The $\Obf_1$ and $\Obf_2$ in (\ref{eq:Obf1}) and (\ref{eq:Obf2}) are
\bqn
\Obf_1=\0bf\quad\textrm{and}\quad\Obf_2=\0bf.
\eqn
Therefore, (\ref{eq:W1O1}) and (\ref{eq:W2O2}) are satisfied. Hence the expressions for $\Sigmabf_1$ and $\Sigmabf_2$ are not relevant. As in Remark \ref{remark:Ricatti}, we only need to show the existence of $\Sigmabf_1$ and $\Sigmabf_2$ that satisfy (\ref{eq:Ei})-(\ref{eq:sigma2}). We have that (\ref{eq:cond}) is also satisfied:
\bqa
\textrm{radius}\left(\Phibf_1\right)=0.4350\quad\textrm{and}\quad\textrm{radius}\left(\Phibf_2\right)=0.3130.\nn
\eqa
Then, all the conditions in Theorem \ref{theorem:MIMO} are satisfied. Therefore, the sum-rate capacity is achieved by treating interference as noise and the optimal input covariances are $\Sbf_1^*$ and $\Sbf_2^*$.
\end{example}
\begin{example}
Consider a MISO IC in the form (\ref{eq:MISOmodelOld}) with channel vectors:
\bqn
&&\hat\hp_1=\left[\begin{array}{c}
   -0.1481\\
   -1.7969\\
    0.1331\\
    0.6644\\
                  \end{array}
\right],\quad \hat\fp_1=\left[\begin{array}{c}
    0.0201\\
   -0.0197\\
   -0.0729\\
    0.7636
                              \end{array}
\right],\quad\hat\hp_2=\left[\begin{array}{c}
    0.1050\\
   -0.0523\\
    1.8070
                               \end{array}
\right],\quad \hat\fp_2=\left[\begin{array}{c}
   -0.4748\\
   -0.7711\\
    0.3813
                              \end{array}
\right]
\eqn
and power constraint
\bqn
\hat P_1=\hat P_2=1.
\eqn
The equivalent MISO IC in the form (\ref{eq:MISOmodelNew}) has channel vectors
\bqn
\hp_1=\left[\begin{array}{c}
    0.3586\\
    0.9335
                        \end{array}\right],
\quad \fp_1=\left[\begin{array}{c}
    0.3985\\
         0
                    \end{array}
\right],
\quad \hp_2=\left[\begin{array}{c}
    0.3818\\
    0.9242
                  \end{array}
\right],
\quad \fp_2=\left[\begin{array}{c}
    0.5426\\
         0
                    \end{array}
\right]
\eqn
and power constraints
\bqn
P_1=3.7100\quad\textrm{and}\quad P_2=3.2789.
\eqn
The corresponding channel parameters are
\bqn
\theta_1=0.3833\pi,\quad\theta_2=0.3753\pi,\quad a_1=0.1588,\quad a_2=0.2944.
\eqn
The optimal input covariance matrices for the equivalent channel are
\bqn
\Sbf_1^*=\left[\begin{array}{cc}
    0.2093&\quad    0.8561\\
    0.8561&\quad    3.5007
               \end{array}
\right]\quad\textrm{and}\quad
\Sbf_2^*=\left[\begin{array}{cc}
    0.1345&\quad    0.6503\\
    0.6503&\quad    3.1445
               \end{array}
\right].
\eqn
The corresponding optimal covariance matrices for the original channel are
\bqn
\hat\Sbf_1^*=\left[\begin{array}{cccc}
    0.0070&\quad    0.0808&\quad   -0.0071&\quad   -0.0187\\
    0.0808&\quad    0.9356&\quad   -0.0820&\quad   -0.2168\\
   -0.0071&\quad   -0.0820&\quad    0.0072&\quad    0.0190\\
   -0.0187&\quad   -0.2168&\quad    0.0190&\quad    0.0502
               \end{array}
\right]\quad\textrm{and}\quad
\hat\Sbf_2^*=\left[\begin{array}{ccc}
    0.0253&\quad    0.0204&\quad    0.1558\\
    0.0204&\quad    0.0164&\quad    0.1253\\
    0.1558&\quad    0.1253&\quad    0.9583
               \end{array}
\right].
\eqn
The $\Gbf_1$, $\Gbf_2$ in (\ref{eq:lowerKKTS1}) and (\ref{eq:lowerKKTS2}) and the Lagrangian multipliers are
\bqn
&&\Gbf_1=\left[\begin{array}{cc}
    0.0442&\quad   -0.0357\\
   -0.0357&\quad   -0.0929
             \end{array}
\right],\quad
\Gbf_2=\left[\begin{array}{cc}
    0.0929&\quad   -0.0420\\
   -0.0420&\quad   -0.1017
             \end{array}
\right]\\
&&\Wbf_1=\left[\begin{array}{cc}
    0.1459&\quad   -0.0357\\
   -0.0357&\quad    0.0087
             \end{array}
\right],\quad
\Wbf_2=\left[\begin{array}{cc}
    0.2033&\quad   -0.0420\\
   -0.0420&\quad    0.0087
             \end{array}
\right]\\
&&\lambda_1= 0.1016,\quad\lambda_2= 0.1104.
\eqn
It can be easily verified that the KKT conditions in (\ref{eq:lowerKKTS1})-(\ref{eq:lowerIneq}) are satisfied.

The $A_1$ and $A_2$ that satisfy (\ref{eq:MCmiso1}) and (\ref{eq:MCmiso2}) (or (\ref{eq:Markov1}) and (\ref{eq:Markov2})) are
\bqn
A_1=0.0992\quad\textrm{and}\quad A_2=0.1156,
\eqn
and the $\sigma_i^2$ and $\bar\sigma_i^2$ in (\ref{eq:sigma1MISO})-(\ref{eq:barSigma2MISO}) are
\bqn
&&\sigma_1^2=0.9874>\bar\sigma_1^2= 0.6277\\
&&\sigma_2^2=0.9891>\bar\sigma_2^2=0.4643.
\eqn
Therefore, by Theorem \ref{theorem:MISO}, the sum-rate capacity of this MISO channel is achieved by treating interference as noise.

We can also verify condition (\ref{eq:W1misoExp1}) with
\bqn
k_1=1.0994 \quad\textrm{and}\quad k_2=0.8133.
\eqn
The $\Obf_1$ and $\Obf_2$ matrices in (\ref{eq:Obf1}) and (\ref{eq:Obf2}) are
\bqn
\Obf_1=\left[\begin{array}{cc}
    0.0679&\quad   -0.0166\\
   -0.0166&\quad    0.0041
             \end{array}
\right]\quad\textrm{and}\quad
\Obf_2=\left[\begin{array}{cc}
    0.1280&\quad   -0.0265\\
   -0.0265&\quad    0.0055
             \end{array}
\right].
\eqn
Since $\Wbf_i\succeq\Obf_i$, by Theorem \ref{theorem:MIMO}, the sum-rate capacity of this MISO channel is achieved by treating interference as noise.

The sum-rate capacity is
\bqn
R_1+R_2= 0.7533+0.7009=1.4543.
\eqn
\end{example}
\begin{example}
Consider a SIMO IC with channel vectors:
\bqn
\hat\hp_1=\left[\begin{array}{c}
   -1.8356\\
    0.0668\\
    0.0355
            \end{array}
\right],\quad
\hat\fp_1=\left[\begin{array}{c}
    1.1136\\
   -0.0346\\
   -0.2537\\
    0.1179
            \end{array}
\right],\quad
\hat\hp_2=\left[\begin{array}{c}
    0.2458\\
    0.0700\\
   -0.6086\\
   -1.2226
            \end{array}
\right],\quad
\hat\fp_2=\left[\begin{array}{c}
    0.1583\\
   -0.6714\\
   -0.5161
            \end{array}
\right]
\eqn
and power constraint
\bqn
P_1=P_2=1.
\eqn
The equivalent SIMO IC is
\bqn
\hp_1=\left[\begin{array}{c}
   -0.2234\\
    0.9747
            \end{array}
\right],\quad
\fp_1=\left[\begin{array}{c}
    0.6252\\
         0
            \end{array}
\right],\quad
\hp_2=\left[\begin{array}{c}
    0.1764\\
    0.9843
            \end{array}
\right],\quad
\fp_2=\left[\begin{array}{c}
    0.6201\\
         0
            \end{array}
\right]
\eqn
with power constraint
\bqn
P_1=3.3753\quad\textrm{and}\quad P_2=1.9304.
\eqn
The corresponding channel parameters are
\bqn
\varphi_1=0.5717\pi,\quad \varphi_2=0.4436\pi,\quad a_1=0.3909,\quad a_2= 0.3845.
\eqn
We simply choose matrices $\Abf_1$ and $\Abf_2$ as in (\ref{eq:simoA1simple}) and (\ref{eq:simoA2simple}):
\bqn
\Abf_1=\left[\begin{array}{cc}
   -0.2434 &\quad        0\\
    0.6094 &\quad        0
             \end{array}
\right]\quad\textrm{and}\quad
\Abf_2=\left[\begin{array}{cc}
    0.2537&\quad         0\\
    0.6103&\quad         0
             \end{array}
\right].
\eqn
We have $\Ibf-\Abf_1^T\Abf_1-\Abf_2\Abf_2^T\succeq\0bf$, $\Ibf-\Abf_1\Abf_1^T-\Abf_2^T\Abf_2\succeq\0bf$ and %
\bqn
\textrm{radius}\left(\Phibf_1\right)=0.2784\quad\textrm{and}\quad\textrm{radius}\left(\Phibf_1\right)=0.2815.
\eqn
Therefore, by Theorem \ref{theorem:SIMO} treating interference as noise achieves the sum-rate capacity and
\bqn
R_1+R_2=0.7297+0.5317=1.2614.
\eqn
We can also use Theorem \ref{theorem:MIMO} to verify the result. The $\Abf_1$ and $\Abf_2$ satisfy (\ref{eq:Markov1}) and (\ref{eq:Markov2}). The numerical radius condition guarantees the existence of $\Sigmabf_1$ and $\Sigmabf_2$ to satisfy (\ref{eq:Ei})-(\ref{eq:sigma2}). Furthermore, we have $W_1=W_2=O_1=O_2=0$. Therefore, all the conditions in Theorem \ref{theorem:MIMO} are satisfied.
\end{example}

\begin{figure*}[htp]
\centerline{\leavevmode \epsfxsize=5.5in \epsfysize=4.45in
\epsfbox{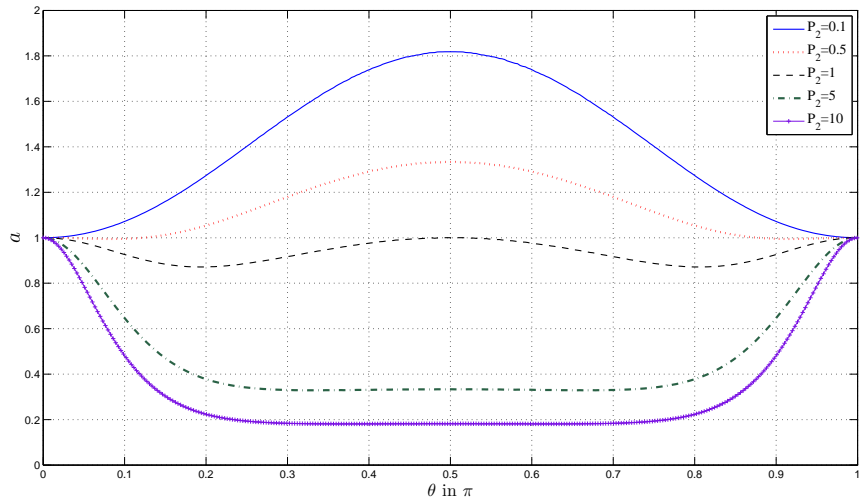}}\caption{The maximum value of $a$ for a MISO ZIC with $P_1=1$ to have noisy interference.}
\label{fig:MISOZ}
\end{figure*}

\begin{example}
In this example, we consider the maximum value of $a_i$ for MISO and SIMO ICs to have noisy interference with various choices of $P_i$ and $\theta_i$ or $\varphi_i$. For the symmetric MISO or SIMO IC, one can use Theorem \ref{theorem:MISO} and \ref{theorem:SIMO} to generate the same result as \cite[Fig. 2]{Annapureddy&Veeravalli:09IT_submission}. For the SIMO ZICs, the maximum $a_2$ is $1$ regardless of $P_i$ and $\varphi_2$ by Proposition \ref{prop:SIMOZ}. For the MISO ZIC, the maximum $a_2$ is shown in Fig. \ref{fig:MISOZ} by Proposition \ref{prop:MISOZ}.
\end{example}
\begin{example}
In this example, we show that a MISO ZIC in which the noisy-interference conditions in Proposition \ref{prop:MISOZ} are violated and treating interference as noise does not achieve the sum-rate capacity.

Consider a MISO ZIC with $P_1=1$, $P_2=10$, $a_1=0$, $a_2=0.4$, $\theta_1=\frac{\pi}{2}$ and $\theta_2=\frac{\pi}{4}$. As is shown in Fig. \ref{fig:MISOZ}, this MISO IC does not satisfy the noisy-interference condition.  The maximum sum-rate by treating interference as noisy is
\bqa
R_1+R_2=1.3725\nn
\eqa
and is achieved by (\ref{eq:MISObfs}) and (\ref{eq:MISOSstar}):
\bqa
\Sbf_1^*=\left[\begin{array}{cc}
                 1 &\quad 0 \\
                 0 &\quad 0
               \end{array}
\right]\quad\textrm{and}\quad \Sbf_2^*=\left[\begin{array}{cc}
    1.7566 &\quad   3.8053\\
    3.8053 &\quad   8.2434
               \end{array}
\right].\nn
\eqa
However, we consider a Han and Kobayashi achievable rate region \cite{Han&Kobayashi:81IT,Chong-etal:08IT} for the MISO ZIC:
\bqa
R_1&{}\leq{}&\frac{1}{2}\log\left(1+\frac{P_1}{1+\fp_2^T\Sbf_p\fp_2}\right)\nn\\
R_2&{}\leq{}&\frac{1}{2}\log\left(1+\hp_2^T\left(\Sbf_p+\Sbf_c\right)\hp_2\right)\nn\\
R_1+R_2&{}\leq{}&\frac{1}{2}\log\left(1+\hp_2^T\Sbf_p\hp_2\right)+\frac{1}{2}\log\left(1+\frac{P_1+\fp_2^T\Sbf_c\fp_2}{1+\fp_2^T\Sbf_p\fp_2}\right)\nn
\eqa
where $\Sbf_p$ and $\Sbf_c$ are respectively the covariance matrices for the input vectors that carry the private and common messages. Then we can achieve a sum-rate of
\bqa
R_1+R_2= 1.4093\nn
\eqa
by the same $\Sbf_1^*$ and a different $\Sbf_2^*=\Sbf_p^*+\Sbf_c^*$ with
\bqa
 \Sbf_p^*=\left[\begin{array}{cc}
    1.1542 &\quad   2.2652\\
    2.2652 &\quad   4.4458
               \end{array}
\right]\quad\textrm{and}\quad
 \Sbf_c^*=\left[\begin{array}{cc}
    4.1906 &\quad   0.9367\\
    0.9367 &\quad   0.2094
               \end{array}
\right].\nn
\eqa
\end{example}

\section{Conclusion}
\label{section:conclusion}

We have studied the noisy-interference sum-rate capacity of MIMO ICs. Sufficient conditions for a MIMO IC to achieve the sum-rate capacity by treating interference as noise have been obtained. For the special cases of MISO and SIMO ICs, simplified conditions have been derived. These conditions largely extend all the existing sufficient conditions.

\appendix
\subsection{Proof of Lemma \ref{lemma:KKT}}
\label{appendix:kkt}
If we write the optimization problem in the standard form:
\bqa
\min&&\quad f\left(\xp\right)\nn\\
\textrm{subject to}&&\quad g_i\left(\xp\right)\leq 0,\quad i=1,\cdots,m\nn\\
&&\quad \xp\in\Xmat
\label{eq:standard}
\eqa
then CQ5 in \cite[p. 306]{Bertsekas-etal:book} requires that there exist a vector $\yp\in N_\Xmat\left(\xp^*\right)^*$ such that
\bqa
\bigtriangledown g_j\left(\xp^*\right)^T\yp<0 \quad \forall j\in A\left(\xp^*\right)
\eqa
where $\xp^*$ is optimal for problem (\ref{eq:standard}), $\bigtriangledown g_j\left(\xp^*\right)$ is the gradient of $g_j(\xp)$ at $\xp^*$, $N_\Xmat\left(\xp^*\right)$ is the normal cone of $\Xmat$ at $\xp^*$, $N_\Xmat\left(\xp^*\right)^*$ is the polar cone of $N_\Xmat\left(\xp^*\right)$, and $A\left(\xp^*\right)$ is index set of all the active inequality constraints. Applying this theorem to our case, we need to find matrices $\Kbf_i$, $i=1,2$, such that
\bqa
&&\Kbf_i\in N_{\Smat_i}\left(\Sbf_i^*\right)^*=T_{\Smat_i}\left(\Sbf_i^*\right)
\label{eq:cq1}\\
&&\tr\left(\Kbf_i\right)<0 \quad \textrm{if } \tr\left(\Sbf_i^*\right)=P_i
\label{eq:cq2}
\eqa
where $\Smat_i$ is the set of symmetric positive semi-definite matrices with the same dimension as that of $\Sbf_i^*$, and $T_{\Smat_i}\left(\Sbf_i^*\right)$ is the tangent cone of $\Smat_i$ at $\Sbf_i^*$. The equality of (\ref{eq:cq1}) is due to the convexity of $\Smat_i$ and \cite[Proposition 4.6.3, p. 254]{Bertsekas-etal:book}.

Define a sequence of matrices $\left\{\Ybf_k\right\}$:
\bqa
\Ybf_k=\Sbf_1^*-\frac{1}{k}\Ubf\cdot\diag[\eta_1,0,\cdots, 0]\cdot\Ubf^T,\quad k=1,2\cdots\cdots
\eqa
where $\Ubf$ is a unitary matrix associated with the eigenvalue decomposition of $\Sbf_1^*$, and $\eta_1$ is the largest eigenvalue of $\Sbf_i^*$:
\bqa
\Sbf_i^*=\Ubf\cdot\diag[\eta_1,\eta_2,\cdots,\eta_{t_i}]\cdot\Ubf^T.
\eqa
Obviously, we have
\bqa
&&\{\Ybf_k\}\subseteq\Smat_1,\quad \Ybf_k\neq\Sbf_1^*\\
&&\lim_{k\rightarrow\infty}\Ybf_k=\Sbf_1^*\\
&&\lim_{k\rightarrow\infty}\frac{\Ybf_k-\Sbf_1^*}{\|\Vec\left(\Ybf_k-\Sbf_1^*\right)\|}=\frac{-\Ubf\cdot\diag[\eta_1,0,\cdots, 0]\cdot\Ubf^T}{\left\|\Vec\left(\Ubf\cdot\diag[\eta_1,0,\cdots, 0]\cdot\Ubf^T\right)\right\|}.
\eqa
Therefore, by \cite[Definition 4.6.2, p. 248]{Bertsekas-etal:book}
\bqa
&&\Kbf_1\triangleq-\Ubf\cdot\diag[\eta_1,0,\cdots, 0]\cdot\Ubf^T\in T_{\Smat_i}\left(\Sbf_i^*\right).
\eqa
Since $\eta_1$ is the largest eigenvalue of $\Sbf_1^*$, we have
\bqa
&&\tr\left(\Kbf_1\right)=-\eta_1<0 \quad\textrm{if }\tr(\Sbf_1^*)=P_1>0.
\eqa
We can similarly find $\Kbf_2$ satisfying (\ref{eq:cq1}) and (\ref{eq:cq2}) for $\Sbf_2^*$. Therefore, the constraint qualifications are satisfied and there exist Lagrangian multipliers $\lambda_i$ and $\Wbf_i$ satisfying (\ref{eq:lowerKKTS1})-(\ref{eq:lowerIneq}).

\subsection{Proof of Lemma \ref{lemma:cvxOpt}}
\label{appendix:cvx}

To prove that the objective function of problem (\ref{eq:upper}) is concave over $\Sbf_1$ and $\Sbf_2$, it is equivalent to prove that (\ref{eq:concaveExp}) is concave. By \cite[Lemma 1]{Shang-etal:10IT_mimo}, both the conditional entropies $h\left(\Hbf_1\xp_{1G}+\Fbf_2\xp_{2G}+\zp_1|\Fbf_1\xp_{1G}+\np_1\right)$ and $h\left(\Hbf_2\xp_{2G}+\Fbf_1\xp_{1G}+\zp_2|\Fbf_2\xp_{2G}+\np_2\right)$ are concave. Therefore, by symmetry, it suffices to prove that $ h\left(\Fbf_1\xp_{1G}+\np_1\right)-h\left(\Fbf_1\xp_{1G}+\zp_2|\np_2\right)$ is concave over $\Sbf_1$ and $\Sbf_2$.

From (\ref{eq:ZN}) we have $\Cov\left(\zp_2|\np_2\right)=\Ibf-\Abf_2\Sigmabf_2^{-1}\Abf_2^T$. From (\ref{eq:sigma1}), there exists a Gaussian vector $\vp\sim\Nmat\left(\0bf,\tilde\Sigmabf\right)$ where
\bqa
\tilde\Sigmabf=\left(\Ibf-\Abf_2\Sigmabf_2^{-1}\Abf_2^T\right)-\Sigmabf_1.\nn
\eqa
We further let $\tilde\zp$ be independent of all other random vectors of interest, and then we have
\bqa
h\left(\Fbf_1\xp_{1G}+\np_1\right)-h\left(\Fbf_1\xp_{1G}+\zp_2|\np_2\right)&{}={}&h\left(\Fbf_1\xp_{1G}+\np_1\right)-h\left(\Fbf_1\xp_{1G}+\np_1+\vp\right)\nn\\
&{}={}&-I\left(\vp;\Fbf_1\xp_{1G}+\np_1+\vp\right).
\eqa
Define a binary random variable $Q$ with probability mass function $Pr(Q=0)=q$ and $Pr(Q=1)=1-q$ where $0\leq q\leq 1$. Let $\bar\xp_1$ have mixed Gaussian distribution with conditional distribution
\bqa
p\left(\bar\xp_1|Q\right)=\left\{\begin{array}{l}
                                    p\left(\bar\xp_1|Q=0\right)=p\left(\bar\xp_1^{(1)}\right)\sim\Nmat\left(\0bf,\Sbf_1^{(1)}\right)\\
                                    p\left(\bar\xp_1|Q=1\right)=p\left(\bar\xp_1^{(2)}\right)\sim\Nmat\left(\0bf,\Sbf_1^{(2)}\right)
                                  \end{array}
\right.
\eqa
where
\bqa
\Sbf_1=q\Sbf_1^{(1)}+(1-q)\Sbf_1^{(2)}.
\label{eq:convexComb}
\eqa
Then we have
\bqa
&&-qI\left(\vp;\Fbf_1\xp_1^{(1)}+\np_1+\vp\right)-(1-q)I\left(\vp;\Fbf_1\xp_2^{(2)}+\np_1+\vp\right)\nn\\
&&=-I\left(\vp;\Fbf_1\bar\xp_1+\np_1+\vp|Q\right)\nn\\
&&=-h\left(\vp|Q\right)+h\left(\vp|\Fbf_1\bar\xp_1+\np_1+\vp,Q\right)\nn\\
&&\stackrel{(a)}\leq-I\left(\vp;\Fbf_1\bar\xp_1+\np_1+\vp\right)\nn\\
&&\stackrel{(b)}\leq -I\left(\vp;\Fbf_1\xp_{1G}+\np_1+\vp\right)
\eqa
where (a) is by the assumption that $Q$ is independent of $\vp$ and the fact that conditioning does not increase entropy. In (b), we let $\xp_{1G}\sim\Nmat\left(\0bf,\Sbf_1\right)$. The inequality is by (\ref{eq:convexComb}) and the fact that Gaussian noise is the worst additive noise \cite{Diggavi&Cover:01IT}. Therefore, $h\left(\Fbf_1\xp_{1G}+\np_1\right)-h\left(\Fbf_1\xp_{1G}+\zp_2|\np_2\right)$ is concave over $\Sbf_1$ and $\Sbf_2$. Similarly, we can prove that $h\left(\Fbf_2\xp_{2G}+\np_2\right)-h\left(\Fbf_2\xp_{2G}+\zp_1|\np_1\right)$ is also a concave function of $\Sbf_1$ and $\Sbf_2$.

\subsection{Proof of Lemma \ref{lemma:myMxInv}}
\label{appendix:mxEq}

In the proof of \cite[Lemma 9]{Shang-etal:10IT_mimo}, if $\textrm{radius}(\Phibf_1)\leq\frac{1}{2}$, then there exist $\Sigmabf_1$ that satisfy
\bqa
\Sigmabf_1&{}={}&\Ibf-\Abf_1\left(\Ibf-\Abf_1\Sigmabf_1^{-1}\Abf_1^T\right)^{-1}\Abf_2
\label{eq:mxEqSigma1}
\eqa
and $\Sigmabf_1-\Abf_1^T\Abf_1$ is positive definite.
Then it suffice to prove that $\Ibf-\Abf_1\Sigmabf_1^{-1}\Abf_1^T$ is positive definite since we can substitute $\Sigmabf_1$ defined in (\ref{eq:mxEqSigma1}) into (\ref{eq:RicattiEq2}) and obtain a positive definite $\Sigmabf_2$.

Let $\Sigmabf_1=\Abf_1^T\Abf_1+\Xbf$ where $\Xbf\succ\0bf$; then we have
\bqa
\Sigmabf_2&{}={}&\Ibf-\Abf_1\Sigmabf_1^{-1}\Abf_1^T\nn\\
&{}={}&\Ibf-\Abf_1\left(\Xbf+\Abf_1^T\Abf_1\right)^{-1}\Abf_1^T\nn\\
&{}\stackrel{(a)}={}&\Ibf-\Abf_1\left(\Xbf+\Tbf\Lambdabf\Tbf^T\right)^{-1}\Abf_1^T\nn\\
&{}\stackrel{(b)}\succeq{}&\Ibf-\Abf_1\left(\eta\Ibf+\Tbf\Lambdabf\Tbf^T\right)^{-1}\Abf_1^T\nn\\
&{}={}&\Ibf-\Abf_1\Tbf\left(\eta\Ibf+\Lambdabf\right)^{-1}\Tbf^T\Abf_1^T
\eqa
where in (a) we let $\Abf_1^T\Abf_1=\Tbf\Lambdabf\Tbf^T$ be the eigenvalue decomposition of $\Abf_1^T\Abf_1$ and $\Tbf\Tbf^T=\Ibf$ and $\Lambdabf$ is a diagonal matrix with non-negative diagonal elements. In (b), we let $\eta$ be the smallest eigenvalue of $\Xbf$. Since $\Xbf$ is symmetric positive definite, we have $\eta>0$. The inequality of (b) is by the fact $\Xbf\succeq\eta\Ibf$.

Since $\Ibf-\Bbf^T\Bbf$ is positive definite if and only if $\Ibf-\Bbf\Bbf^T$ is positive definite, we only need to prove that $\Ibf-\left(\eta\Ibf+\Lambdabf\right)^{-\frac{1}{2}}\Tbf^T\Abf_1^T\Abf_1\Tbf\left(\eta\Ibf+\Lambdabf\right)^{-\frac{1}{2}}$ is positive definite, which is obviously true since $\Tbf^T\Abf_1^T\Abf_1\Tbf=\Lambdabf$ and $\eta>0$.

We have proved that if $\textrm{radius}\left(\Phibf_1\right)\leq\frac{1}{2}$, then there exist $\Sigmabf_1\succ\Abf_1^T\Abf_1$ and $\Sigmabf_2\succ\0bf$ that satisfy (\ref{eq:RicattiEq1}) and (\ref{eq:RicattiEq2}). Now we need to prove that $\Sigmabf_2\succ\Abf_2^T\Abf_2$, which is true by the fact $\Ibf-\Abf_2\Sigmabf_2^{-1}\Abf_2^T=\Sigmabf_1\succ\0bf$ and \cite[Lemma 6]{Shang-etal:10IT_mimo}.

By symmetry, if $\textrm{radius}(\Phibf_2)\leq\frac{1}{2}$, we also have positive definite solutions. The equivalence between (\ref{eq:SigmabfAA}) and (\ref{eq:COVfeasible}) is by \cite[Lemma 6]{Shang-etal:10IT_mimo}.

\bibliography{Journal,Conf,Misc,Book}

\begin{thebibliography}{10}

\bibitem{Shannon:61Berkeley}
C.~E. Shannon,
\newblock ``{Two-way communication channels},''
\newblock in {\em Proc. 4th Berkeley Symp. Math. Stat. and Prob.}, Berkeley,
  CA, 1961, vol.~1, pp. 611--644,
\newblock Also available in {\em Claude E. Shannon: Collected Papers}, IEEE
  Press, New York, 1993.

\bibitem{Ahlswede:74AP}
R.~Ahlswede,
\newblock ``{The capacity region of a channel with two senders and two
  receivers},''
\newblock {\em Ann. Prob.}, vol. 2, no. 5, pp. 805--814, 1974.

\bibitem{Carleial:75IT}
A.~B. Carleial,
\newblock ``{A case where interference does not reduce capacity},''
\newblock {\em IEEE Trans. Inf. Theory}, vol. 21, pp. 569--570, Sep. 1975.

\bibitem{Sato&Tanabe:78IECE}
H.~Sato and M.~Tanabe,
\newblock ``{A discrete two-user channel with strong interferene},''
\newblock {\em Trans. IECE Japan}, vol. 61, pp. 880--884, 1978.

\bibitem{Carleial:78IT}
A.~B. Carleial,
\newblock ``{Interference channels},''
\newblock {\em IEEE Trans. Inf. Theory}, vol. 24, pp. 60--70, Jan. 1978.

\bibitem{Han&Kobayashi:81IT}
T.~S. Han and K.~Kobayashi,
\newblock ``{A new achievable rate region for the interference channel},''
\newblock {\em IEEE Trans. Inf. Theory}, vol. 27, pp. 49--60, Jan. 1981.

\bibitem{Chong-etal:08IT}
H.~F. Chong, M.~Motani, H.~K. Garg, and H.~El Gamal,
\newblock ``{On the Han-Kobayashi region for the interference channel},''
\newblock {\em IEEE Trans. Inf. Theory}, vol. 54, pp. 3188--3195, Jul. 2008.

\bibitem{Kramer:06Zurich}
G.~Kramer,
\newblock ``{Review of rate regions for interference channels},''
\newblock in {\em {Proc. International Zurich Seminar}}, Zurich, Switzerland,
  Feb. 2006, pp. 162--165.

\bibitem{Sato:77IT}
H.~Sato,
\newblock ``{Two-user communication channels},''
\newblock {\em IEEE Trans. Inf. Theory}, vol. 23, pp. 295--304, May 1977.

\bibitem{Sato:78IT}
H.~Sato,
\newblock ``{On degraded Gaussian two-user channels},''
\newblock {\em IEEE Trans. Inf. Theory}, vol. 24, pp. 634--640, Sep. 1978.

\bibitem{Carleial:83IT}
A.~B. Carleial,
\newblock ``{Outer bounds on the capacity of interference channels},''
\newblock {\em IEEE Trans. Inf. Theory}, vol. 29, pp. 602--606, July 1983.

\bibitem{Sato:81IT}
H.~Sato,
\newblock ``{The capacity of the Gaussian interference channel under strong
  interference},''
\newblock {\em IEEE Trans. Inf. Theory}, vol. 27, pp. 786--788, Nov. 1981.

\bibitem{Costa&ElGamal:87IT}
M.~H.~M. Costa and A.~A.~El Gamal,
\newblock ``{The capacity region of the discrete memoryless interference
  channel with strong interference},''
\newblock {\em IEEE Trans. Inf. Theory}, vol. 33, pp. 710--711, Sep. 1987.

\bibitem{Benzel:79IT}
R.~Benzel,
\newblock ``{The capacity region of a class of discrete additive degraded
  interference channels},''
\newblock {\em IEEE Trans. Inf. Theory}, vol. 25, pp. 228--231, Mar. 1979.

\bibitem{Liu&Ulukus:08IT}
N.~Liu and S.~Ulukus,
\newblock ``{The capacity region of a class of discrete degraded interference
  channels},''
\newblock {\em IEEE Trans. Inf. Theory}, vol. 54, pp. 4372--4378, Sep. 2008.

\bibitem{Costa:85IT}
M.~H.~M. Costa,
\newblock ``{On the Gaussian interference channel},''
\newblock {\em IEEE Trans. Inf. Theory}, vol. 31, pp. 607--615, Sept. 1985.

\bibitem{Sason:04IT}
I.~Sason,
\newblock ``{On achievable rate regions for the Gaussian interference
  channels},''
\newblock {\em IEEE Trans. Inf. Theory}, vol. 50, pp. 1345--1356, June 2004.

\bibitem{Cheng&Verdu:93IT}
R.~S. Cheng and S.~Verd\'{u},
\newblock ``{On limiting characterizations of memoryless multiuser capacity
  regions},''
\newblock {\em Trans. Inf. Theory}, vol. 39, pp. 609--612, 1993.

\bibitem{Kramer:04IT}
G.~Kramer,
\newblock ``{Outer bounds on the capacity of Gaussian interference channels},''
\newblock {\em IEEE Trans. Inf. Theory}, vol. 50, pp. 581--586, Mar. 2004.

\bibitem{Etkin-etal:08IT}
R.~H. Etkin, D.~Tse, and H.~Wang,
\newblock ``{Gaussian interference channel capacity to within one bit},''
\newblock {\em IEEE Trans. Inf. Theory}, vol. 54, no. 12, pp. 5534--5562, Dec.
  2008.

\bibitem{Shang-etal:09IT}
X.~Shang, G.~Kramer, and B.~Chen,
\newblock ``{A new outer bound and the noisy-interference sum-rate capacity for
  Gaussian interference channels},''
\newblock {\em IEEE Trans. Inf. Theory}, vol. 55, no. 2, pp. 689--699, Feb.
  2009.

\bibitem{Motahari&Khandani:09IT}
A.~S. Motahari and A.~K. Khandani,
\newblock ``{Capacity bounds for the Gaussian interference channel},''
\newblock {\em IEEE Trans. Inf. Theory}, vol. 55, no. 2, pp. 620--643, Feb.
  2009.

\bibitem{Annapureddy&Veeravalli:09IT}
V.~S. Annapureddy and V.~V. Veeravalli,
\newblock ``{Gaussian interference networks: Sum capacity in the low
  interference regime and new outer bounds on the capacity region},''
\newblock {\em IEEE Trans. Inf. Theory}, vol. 55, no. 7, pp. 3032--3050, Jul.
  2009.

\bibitem{Shang-etal:08ISIT}
X.~Shang, G.~Kramer, and B.~Chen,
\newblock ``{New outer bounds on the capacity region of Gaussian interference
  channels},''
\newblock in {\em Proc. IEEE International Symposium on Information Theory},
  Toronto, Canada, Jul. 2008, pp. 245 -- 249.

\bibitem{Shang-etal:08Milcom}
X.~Shang, G.~Kramer, and B.~Chen,
\newblock ``{Throughput optimization for multi-user interference channels},''
\newblock in {\em Proc. IEEE Military Communications Conference}, San Diego,
  CA, Nov. 2008, pp. 1--7.

\bibitem{Weng&Tuninetti:08ITA}
Y.~Weng and D.~Tuninetti,
\newblock ``{On Gaussian interference channels with mixed interference},''
\newblock in {\em Proc. Information Theory and Applications Workshop}, San
  Diego, CA, Jan. 2008,
\newblock [Online]. Available:
  http://ita.ucsa/workshop/08/files/paper/paper\_264.pdf.

\bibitem{Vishwanath&Jafar:04ITW}
S.~Vishwanath and S.~A. Jafar,
\newblock ``{On the capacity of vector Gaussian interference channels},''
\newblock in {\em Proc. IEEE Information Theory Workshop}, San Antonio, TX,
  Oct. 2004, pp. 365--369.

\bibitem{Shang-etal:06IT}
X.~Shang, B.~Chen, and M.~J. Gans,
\newblock ``{On achievable sum rate for MIMO interference channels},''
\newblock {\em IEEE Trans. Inf. Theory}, vol. 52, no. 9, pp. 4313--4320, Sep.
  2006.

\bibitem{Telatar&Tse:07ISIT}
E.~Telatar and D.~Tse,
\newblock ``{Bounds on the capacity region of a class of interference
  channels},''
\newblock in {\em {Proc. IEEE International Symposium on Information Theory}},
  Nice, France, Jun. 2007, pp. 2871--2874.

\bibitem{Shang-etal:08Allerton}
X.~Shang, B.~Chen, G.~Kramer, and H.~V. Poor,
\newblock ``{On the capacity of MIMO interference channels},''
\newblock in {\em Proc. $46$th Annual Allerton Conference on Communication,
  Control, and Computing}, Monticello, IL, Sep. 2008, pp. 700--707.

\bibitem{Shang-etal:10IT_mimo}
X.~Shang, B.~Chen, G.~Kramer, and H.~V. Poor,
\newblock ``{Capacity regions and sum-rate capacities of vector Gaussian
  interference channels},''
\newblock {\em IEEE Trans. Inf. Theory}, vol. 56, no. 10, pp. 5030--5044, Oct.
  2010.

\bibitem{Annapureddy&Veeravalli:09IT_submission}
V.~S. Annapureddy and V.~V. Veeravalli,
\newblock ``{Sum capacity of MIMO interference channels in the low interference
  regime},''
\newblock {\em {\em submitted to} IEEE Trans. Inf. Theory.
  http://arxiv.org/abs/0909.2074}, Sep. 2009.

\bibitem{Shang-etal:11IT_pgic}
X.~Shang, B.~Chen, G.~Kramer, and H.~V. Poor,
\newblock ``{Noisy-interference sum-rate capacity of parallel Gaussian
  interference channels},''
\newblock {\em IEEE Trans. Inf. Theory}, vol. 57, no. 1, pp. 210--226, Jan.
  2011.

\bibitem{Shang&Chen:07Asi}
X.~Shang and B.~Chen,
\newblock ``{Achievable rate region for downlink beamforming in the presence of
  interference},''
\newblock in {\em {Proc. 41st Asilomar Conference on Signals, Systems, and
  Computers}}, Pacific Grove, CA, Nov. 2007, pp. 1684--1688.

\bibitem{Shang-etal:09IT_submission2}
X.~Shang, B.~Chen, and H.~V. Poor,
\newblock ``{Multi-user MISO interference channels with single-user detection:
  optimality of beamforming and the achievable rate region},''
\newblock {\em IEEE Trans. Inf. Theory, {\em to appear },
  http://arxiv.org/abs/0907.0505}, Apr. 2009.

\bibitem{Zhang&Cui:10SP}
R.~Zhang and S.~Cui,
\newblock ``{Cooperative interference management with MISO beamforming},''
\newblock {\em IEEE Trans. Signal Processing}, vol. 58, pp. 5450--5458, Oct.
  2010.

\bibitem{Mochaourab&Jorswieck:10SP_submission}
R.~Mochaourab and E.~Jorswieck,
\newblock ``{Optimal beamforming in interference networks with perfect local
  channel information},''
\newblock {\em {\em submitted to} IEEE Trans. Signal Processing,
  http://arxiv.org/abs/1004.4492}, Oct. 2010.

\bibitem{Horn&Johnson:book}
R.~A. Horn and C.~R. Johnson,
\newblock {\em Matrix Analysis},
\newblock Cambridge University Press, New York, 1985.

\bibitem{Bertsekas-etal:book}
D.~P. Bertsekas, A.~Nedic, and A.~E. Ozdaglar,
\newblock {\em Convex Analysis and Optimization},
\newblock Athena Scientific, Belmont, MA, 2003.

\bibitem{Searle:book}
S.~R. Searle,
\newblock {\em Matrix Algebra Useful for Statistics},
\newblock John Wiley \& Sons, Inc., New York, 1982.

\bibitem{Bartels&Stewart:72ACM}
R.~H. Bartels and G.~W. Stewart,
\newblock ``{Solution of the matrix equation AX + XB = C},''
\newblock {\em Communications of the ACM}, vol. 15, pp. 820--826, Sep. 1972.

\bibitem{Engwerda-etal:93LA&A}
J.~C. Engwerda, A.~C.~M. Ran, and A.~L. Rijkeboer,
\newblock ``{Necessary and sufficient conditions for the existence of a
  positive definite solution of the matrix equation
  $\Xbf+\Abf^*\Xbf^{-1}\Abf=\Qbf$},''
\newblock {\em Linear Algebra and Its Applications}, vol. 186, pp. 255--275,
  1993.

\bibitem{Diggavi&Cover:01IT}
S.~N. Diggavi and T.~M. Cover,
\newblock ``{The worst additive noise under a covariance constraint},''
\newblock {\em IEEE Trans. Inf. Theory}, vol. 47, no. 7, pp. 3072--3081, Nov.
  2001.

\end{thebibliography}
\bibliographystyle{IEEEbib}
\end{document}